%% file: main.tex
\documentclass[lettersize,journal]{IEEEtran}
\usepackage{amsmath,amsfonts}
\usepackage{algorithm}
\usepackage{array}
\usepackage[caption=false,font=footnotesize,labelfont=rm,textfont=rm]{subfig}
\usepackage{textcomp}
\usepackage{stfloats}
\usepackage{url}
\usepackage{verbatim}
\usepackage{graphicx}
\usepackage{cite}
\usepackage{siunitx}
\usepackage[colorlinks,linkcolor=blue,anchorcolor=blue,citecolor=blue]{hyperref}
\usepackage[dvipsnames]{xcolor}
\usepackage{booktabs}      
\usepackage{multirow}      
\usepackage{makecell}      
\usepackage{siunitx}       
\usepackage{subcaption}
\usepackage{algpseudocode}
\usepackage{subfig}
\usepackage{tabularx}
\usepackage[table]{xcolor}         
\definecolor{OursBlue}{RGB}{220,235,255}
\newcolumntype{Y}{>{\centering\arraybackslash}X}

\begin{document}

\title{YODA: Yet Another One-step Diffusion-based Video Compressor}

\author{Xingchen Li, Junzhe Zhang, Junqi Shi, Ming Lu, and Zhan Ma
\thanks{X. Li, J. Zhang, J. Shi, M. Lu, and Z. Ma are with the School of Electronic Science and Engineering, Nanjing University, Nanjing, Jiangsu 210023, China (e-mail: xingchenli@smail.nju.edu.cn, junzhezhang@smail.nju.edu.cn, junqishi@smail.nju.edu.cn; minglu@nju.edu.cn; mazhan@nju.edu.cn).}
}

\markboth{}{}

\maketitle
\begin{abstract}
While one-step diffusion models have recently excelled in perceptual image compression, their application to video remains limited. Prior efforts typically rely on pretrained 2D autoencoders that generate per-frame latent representations independently, thereby neglecting temporal dependencies. We present YODA—Yet Another One-step Diffusion-based Video Compressor—which embeds multiscale features from temporal references for both latent generation and latent coding to better exploit spatial–temporal correlations for more compact representation, and employs a linear Diffusion Transformer (DiT) for efficient one-step denoising. YODA achieves state-of-the-art perceptual performance, consistently outperforming traditional and deep-learning baselines on LPIPS, DISTS, FID, and KID. Source code will be publicly available at \url{https://github.com/NJUVISION/YODA}.

\end{abstract}


\input{sec/1_intro}

\input{sec/2_related}

\input{sec/3_method}

\input{sec/4_exp}

\input{sec/5_conclusion}

\bibliographystyle{IEEEbib}
\bibliography{main}

\vfill

\end{document}

%% file: sec/1_intro.tex
\section{Introduction}
\label{sec:intro}

\IEEEPARstart{R}{ecent} advances in neural video compression (NVC) have fundamentally reshaped video coding~\cite{chen2017deepcoder,lu2019dvc,liu2020learned,li2021dcvc,dcvcfm,dhvc,DCVCRT}. By optimizing latent representations through data-driven learning, neural codecs now deliver superior rate–distortion (R-D) performance compared with established standards such as H.264/AVC~\cite{264}, H.265/HEVC~\cite{HEVC}, and H.266/VVC~\cite{vvc}. These models exploit spatial–temporal correlations more effectively than traditional designs, achieving substantially lower bitrates while preserving high objective fidelity.

{Despite recent progress, the majority of neural video codecs remain anchored to pixel distortion-oriented optimization inherited from conventional standards, typically targeting metrics like PSNR (peak signal-to-noise ratio). While such objectives favor pixel-level accuracy, they correlate weakly with human perception, particularly at low bitrates where perceptual quality is paramount. This misalignment has drawn attention to perceptually or subjectively optimized NVC, which emphasizes visually pleasing reconstruction by incorporating perceptual losses~\cite{PLVC}, generative models~\cite{GLCvideo}, etc. The overarching aim is a human-aligned rate–quality trade-off in which subjective realism takes precedence over pixel fidelity.}

\begin{figure}[htbp]
    \centering
    \subfloat[]{
        \includegraphics[width=\linewidth]{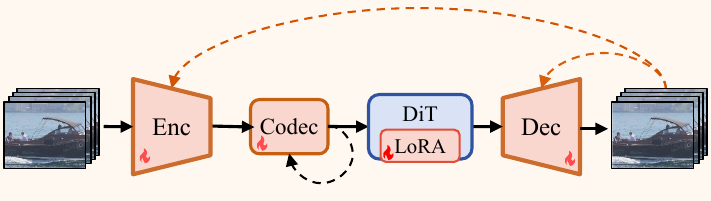} \label{fig:compare_YODA}
    }\hfill    
        \subfloat[]{
        \includegraphics[width=\linewidth]{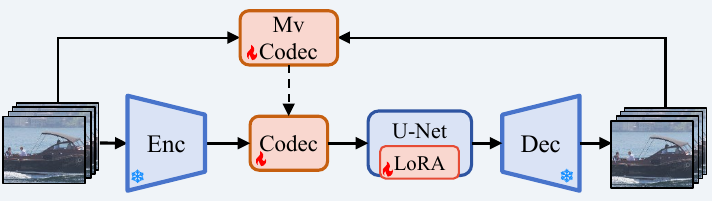}  \label{fig:compare_diffOSD}
    }

        \caption{(a) YODA adopts a trainable temporal-aware autoencoder, a latent codec that models motion implicitly, and a linear DiT-based denoiser; while (b) current approaches use a frozen autoencoder that operates only in the spatial domain, a latent codec that explicitly encodes motion, and a U-Net denoiser. For the diffusion denoiser, LoRA fine-tuning is applied.}
    \label{fig:compare_framework}
\end{figure}

{Motivated by the superior generative abilities of diffusion models~\cite{DDPM, VAEDiff}, researchers have explored their application to perceptual image compression. A prevalent design couples a pretrained variational autoencoder (VAE) or autoencoder (AE) to produce latent representations, which are then refined by a latent diffusion model. Unlike standard multi-step diffusion sampling, which starts from Gaussian noise, recent one-step diffusion–based image codecs initialize denoising directly from the decoded latents \cite{oscar,zhang2025stablecodec}. This warm start preserves semantic content, reduces the number of diffusion steps dramatically, and yields faithful reconstructions with substantially improved perceptual quality.} 

{On the other hand, this paradigm has so far seen only limited extension to video. One notable example is DiffVC-OSD~\cite{Diffvcosd}, a one-step diffusion–based codec that conditions a denoising U-Net on decoded latents and temporal conditions to reconstruct video frames one-by-one. In DiffVC-OSD (see Fig. \ref{fig:compare_diffOSD}), a pretrained autoencoder first extracts latent representations for each frame, which are then entropy-coded by a learned latent coder (the so-called Contextual Coder in \cite{Diffvcosd}). To inject temporal context to support both latent coding and diffusion guidance, the method adopts a hybrid conditional coding architecture inspired by DCVC-DC~\cite{DCVCDC}, in which a dedicated motion codec aligns frames and embeds temporal references for conditional coding.}

{Whereas in DiffVC-OSD (see Fig. \ref{fig:compare_diffOSD}), the pretrained, frozen autoencoder leverages only per-frame spatial correlations, which is inherently suboptimal for producing sufficiently compact latents without incorporating temporal references as conditioning signals. Moreover, introducing an explicit motion codec complicates the system design \cite{DCVCRT}, given that temporal motion can also be modeled probabilistically in an implicit manner~\cite{dhvc}.}

\begin{table}[htbp]
    \centering
    \caption{Notations}
    \label{tab:notations}
    \begin{tabular}{c|c} 
    \hline
     Item    & Description \\
    \hline
     NVC       & Neural Video Compression\\
     VAE  & Variational AutoEncoder\\
     AE & AutoEncoder\\
     GAN & Generative Adversarial Network\\
     DiT & Diffusion Transformer\\
     MSE & Mean Squared Error\\
     PSNR & Peak Signal-to-Noise Ratio\\
     MS-SSIM & Multiscale Structural Similarity\\
     LPIPS & Learned Perceptual Image Patch Similarity\\
     DISTS & Deep Image Structure and Texture Similarity\\
     FID & Fr\'echet Inception Distance\\
     KID & Kernel Inception Distance\\
     LoRA &  Low-Rank Adaptation \\
     \hline
    \end{tabular}
\end{table}

{Building on this gap, this work introduces YODA - Yet another One-step Diffusion-based Video Coder. YODA makes the following novel parts shown in Fig.~\ref{fig:compare_YODA}:}  
\begin{itemize}
    \item We propose a trainable frame autoencoder—departing from the pretrained, frozen autoencoder commonly used in diffusion-based coders—that embeds multiscale temporal features from prior reference frames. 
    \begin{itemize}
        \item  By explicitly leveraging inter-frame conditioning, the frame autoencoder yields a more compact latent representation—typically reducing its size by a half compared to existing approaches.
        \item By jointly embedding spatiotemporal features, the latent representation becomes better aligned with an entropy model that exploits both spatial and temporal contexts in the latent space. This alignment enables implicit, probabilistic characterization of temporal motion without requiring explicit motion processing.
    \end{itemize}
     \item {We expand the channel dimension in the latent codec (\textit{e.g.}, by 8$\times$ from 32 to 256 in this work), enabling richer contextual information across frames to be effectively captured and exploited for conditional coding (i.e., feature embedding and entropy modeling). This enhancement strengthens the model’s ability to model temporal correlations between frames.}
    \item We further replace the U-Net denoiser, which is predominantly used in prior work~\cite {oscar,Diffvcosd,zhang2025stablecodec}, with a lightweight linear DiT for one-step denoising. This architecture maintains effectiveness while markedly reducing computational cost, enabling end-to-end multi-step training on commodity GPUs.
    
\end{itemize}

 Extensive experiments have demonstrated that the proposed YODA delivers a superior rate–quality tradeoff compared with existing methods, including Diffusion-based approaches like DiffVC~\cite{DiffVC} and DiffVC-OSD~\cite{Diffvcosd}, GAN-based solutions like GLC-Video~\cite{GLCvideo} and PLVC~\cite{PLVC}, VAE-based DCVC-RT~\cite{DCVCRT}, as well as the traditional standards H.265/HEVC and H.266/VVC. These results establish YODA as a new benchmark in diffusion-based video compression.  Table \ref{tab:notations} lists frequently used notations throughout the paper.

%% file: sec/2_related.tex
\section{Related Work}

\subsection{Neural Video Compression }

{{\bf Optimization Towards Better Objective Fidelity.}
In recent years, end-to-end neural video compression has grown exponentially. Early approaches largely mimic the hybrid coding paradigm of traditional standards by stacking neural modules under a paired VAE structure for explicit motion estimation/compensation and residual coding. Representative examples include DeepCoder~\cite{chen2017deepcoder}, DVC~\cite{lu2019dvc}, and FVC~\cite{fvc}, etc. Subsequent work explored conditional coding to replace explicit residual coding, leading to a series of advances—e.g., CodecNet~\cite{ladune2020optical} and DCVC variants~\cite{li2021dcvc,DCVCDC,dcvcfm}—while still retaining explicit motion processing. More recently, a line of research implicitly characterizes temporal motion in latent space via probabilistic modeling, offering lower computational cost and a simpler design. Notable efforts include VCT~\cite{mentzer2022vct}, DHVC~\cite{dhvc}, and DCVC-RT~\cite{DCVCRT}.}

The aforementioned methods are primarily trained with mean squared error (MSE), aiming for an optimal trade-off between bitrate and objective reconstruction fidelity (e.g., PSNR). Yet pixel-level losses such as MSE often diverge from human perceptual judgments. As noted in~\cite{tradeoff}, there exists an intrinsic conflict between minimizing pixel distortion and achieving high perceptual quality (realism), making it difficult to optimize both simultaneously, especially at low bitrates.

{\bf Optimization Towards Better Perceptual Realism.}

{To pursue perceptual realism in NVC, a common strategy is to leverage generative models.  Approaches using adversarial loss—such as PLVC~\cite{PLVC} and GLC-Video~\cite{GLCvideo}—use GANs to align the reconstructed distribution with that of natural videos, thereby improving perceptual quality.}

More recently, building on the success of diffusion models in perceptual image compression, diffusion-based methods have been introduced for video compression to further enhance perceptual quality~\cite{videousingdiffusion,DiffVC}. However, multi-step diffusion sampling is computationally prohibitive for practical deployment. To mitigate this, one-step denoising has been explored for efficient inference~\cite{Diffvcosd}.

In one-step diffusion–based video codecs, practitioners typically adopt LoRA fine-tuning with single-step diffusion. This combination substantially reduces inference complexity while improving perceptual quality and often boosting fidelity—because denoising begins from semantically rich compressed latents rather than pure noise.

\subsection{Latent Diffusion Models}

{Diffusion-based generative models have achieved remarkable success in high-fidelity image synthesis. Fundamentally, these models define a parameterized Markov chain to generate samples. The forward process gradually corrupts data with Gaussian noise until it becomes indistinguishable from pure noise, while the reverse process learns to iteratively denoise the signal to reconstruct the original data distribution~\cite{diffusion}. To mitigate the high computational costs of pixel-space diffusion, Latent Diffusion Models (LDM)~\cite{latent_diffusion} incorporate perceptual compression to shift the diffusion process into a lower-dimensional latent space. This paradigm dramatically improves scalability and reduces computational complexity while preserving essential semantic fidelity.}

Building on LDMs, recent work~\cite{dit} replaces convolutional U-Nets with Transformer-based architectures to further boost modeling capacity and scale. For example, Stable Diffusion 3 (SD3)~\cite{sd3} adopts a multimodal DiT-style architecture with modality-specific Transformer blocks for stronger semantic alignment and text rendering, while PixArt-$\alpha$~\cite{chen2023pixart} demonstrates that Transformer-based latent diffusion with a VAE tokenizer can train efficiently while delivering strong high-resolution results.

In parallel, diffusion acceleration has advanced rapidly. On the architectural front, SANA~\cite{xie2025sana} employs a Linear Diffusion Transformer (Linear DiT) to significantly reduce computational complexity. Meanwhile, consistency-based approaches like sCM~\cite{sCM} and LCM~\cite{lcm} enable few-step inference, paving the way for SANA-Sprint~\cite{sanasprint} to achieve high fidelity in just 1--4 steps.

\label{sec:related_work}

%% file: sec/3_method.tex
\section{Method}
\begin{figure*}[htbp]
    \centering
    \includegraphics[width=1\linewidth]{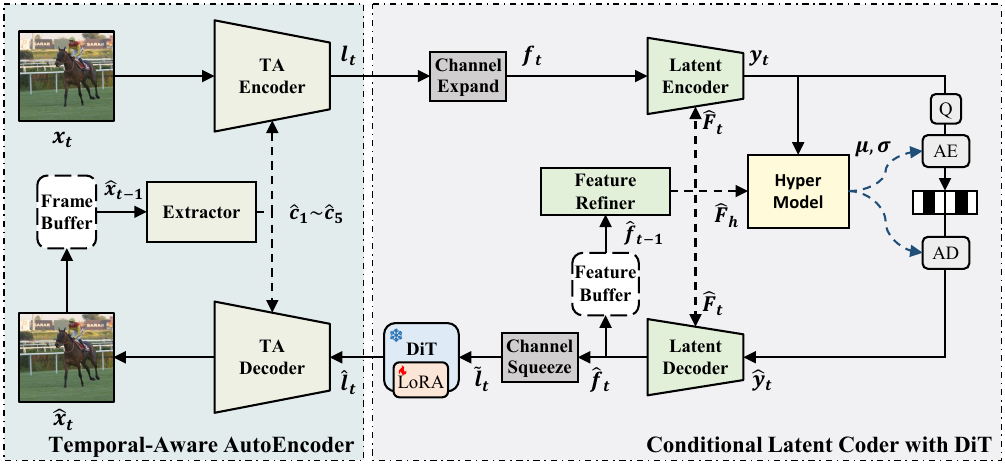}
\caption{{\bf YODA}.  The current frame \(x_t\) is first processed by the Temporal-Aware Encoder (TA Encoder) to produce \(l_t\), and then passed through a channel expansion (Channel-expand) block to obtain \(f_t\), which increases the channel dimensionality from 32 to 256. These features are subsequently compressed under the guidance of a Hyper Model.  In addition to serving as conditions, the decoded features \(\hat{f}_t\) are passed through a channel squeezing (Ch-squeeze) module that reduces the channel dimensionality back to 32, yielding \(\tilde{l}_t\). The representation \(\tilde{l}_t\) is then denoised by a linear DiT module to obtain \(\hat{l}_t\), after which the Temporal-Aware Decoder (TA Decoder) reconstructs the image \(\hat{x}_t\).   An Extractor forms a temporal feedback loop by extracting multiscale cues \( \{\hat{c}_i\}_{i=1}^5\) from the previous reconstruction \(\hat{x}_{t-1}\) and injecting them back into the main encoder–decoder backbone. Q, AE, and AD stand for quantization, arithmetic encoding, and arithmetic decoding, respectively.}

    \label{fig:framework}
\end{figure*}

{YODA comprises three primary components (Fig.~\ref{fig:framework}): a Temporal-Aware AutoEncoder (TA-AE), a Conditional Latent Coder (CLC), and a One-Step DiT Denoiser. Dedicated extractors are designed for the TA-AE and CLC to aggregate cross-frame references tailored for feature formation and entropy modeling. }

Consider an $N$-frame video sequence $\{x_t\}_{t=0}^{N-1}$, where each frame $x_t \in \mathbb{R}^{H\times W\times3}$ has spatial resolution $H\times W$ with three RGB channels (assuming standard 3-channel input). Here, $H$ and $W$ denote the height and width of the video frame, respectively. As illustrated in Fig.~\ref{fig:framework}, given a frame $x_t$, 
\begin{enumerate}
    \item YODA first encodes it into a latent tensor $l_t$ using an (Frame) Encoder of the proposed Temporal-Aware AutoEncoder (TA-AE). A symmetric Decoder then maps the denoised latent $\hat{l}_t$ back to the decoded frame $\hat{x}_t$. To achieve a more compact latent $l_t$, multiscale features from temporal references\footnote{In this work, we use a single reference frame for temporal conditioning.} of $x_t$ are extracted and embedded as conditions, enabling effective exploitation of both spatial and temporal correlations. 
    \item The latent vector $l_t$ is then processed by a Conditional Latent Coder (CLC) —largely following the architecture of DCVC-RT~\cite{DCVCRT}—to produce compressed binary codes. This module aggregates spatial and temporal contexts in latent space to refine probability estimates for entropy coding, thereby improving compression efficiency. In CLC, the channel dimension of internal features is expanded to 256 to better mine the temporal context for information propagation across frames.
\item The DiT model ingests the feature-space latent $\tilde{l}_t$ decoded from the Conditional Latent Coder—now augmented by compression noise—and performs one-step denoising to produce $\hat{l}_t$. This denoised latent is then fed into the TA Decoder to reconstruct $\hat{x}_t$. The DiT module follows the linear DiT structure within the SANA framework~\cite{xie2025sana}, chosen for its efficient training and inference capabilities.
\end{enumerate}

\subsection{Temporal-Aware AutoEncoder (TA-AE)}
\label{sec:temporal_ae}
{Existing diffusion-based video codecs typically reuse a pretrained AE or VAE to produce latents from input frames. This practice has two key limitations: 1) The pretrained autoencoder treats each frame independently, failing to exploit cross-frame dependencies; 2) The latent shape is constrained to $(H/8)\times(W/8)\times4$ to match the dimensionality requirements of subsequent U-Net denoisers, which drives up computational cost and hinders scalability.}

\begin{figure}[t]
    \centering
    \includegraphics[width=1.06\linewidth]{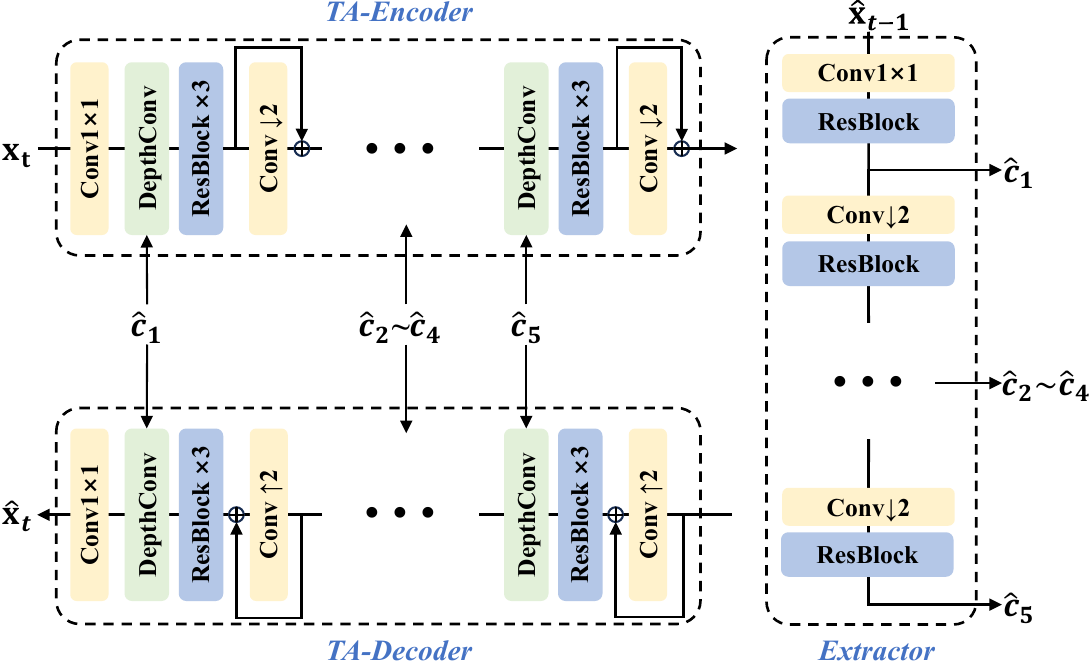}
\caption{{\bf Temporal-Aware Autoencoder (TA-AE)} augments the standard DC‑AE by incorporating multiscale temporal features, $\{\hat{c}_i\}_{i=1}^5$, extracted from the reference frame $\hat{x}_{t-1}$ through the use of Extractor. }    \label{fig:temporalAE}
\end{figure}

{To overcome the limitations of spatial-only encoding, we introduce a temporal-aware autoencoder (TA-AE) that augments SANA’s DC-AE (Deep Compression AutoEncoder) with explicit temporal conditioning (see Fig. \ref{fig:temporalAE}). TA-AE injects multiscale temporal features, i.e., $\{\hat{c}_i\}_{i=1}^5$,  into both the Encoder and Decoder via straightforward concatenation. Specifically, features computed from a temporal reference are integrated at all five spatial resolutions—from $H\times W$ to $(H/16)\times(W/16)$—ensuring that fine-grained temporal priors guide latent generation. A five-scale extractor processes the reconstructed reference frame, $\hat{x}_{t-1}$, to produce these features, i.e., $\{\hat{c}_i\}_{i=1}^5 = {\rm Extractor}\left( \hat{x}_{t-1} \right)$.} 

{The Encoder of YODA's TA-AE maps each input frame  $x_t$ into a latent tensor  $l_t \in \mathbb{R}^{(H/32)\times(W/32)\times d_l}$.  The channel dimension $d_l$ is set to 32, aligning with the DiT denoiser’s expected input. This yields a latent vector size of $(H\times W)/32$-half the resolution used in \cite{Diffvcosd}, which operates at  $(H\times W)/16$. }

\subsection{Conditional Latent Coder (CLC)}
\label{sec:conditional_latent_codec}

\begin{figure}[t]
    \centering
    \includegraphics[width=0.98\linewidth]{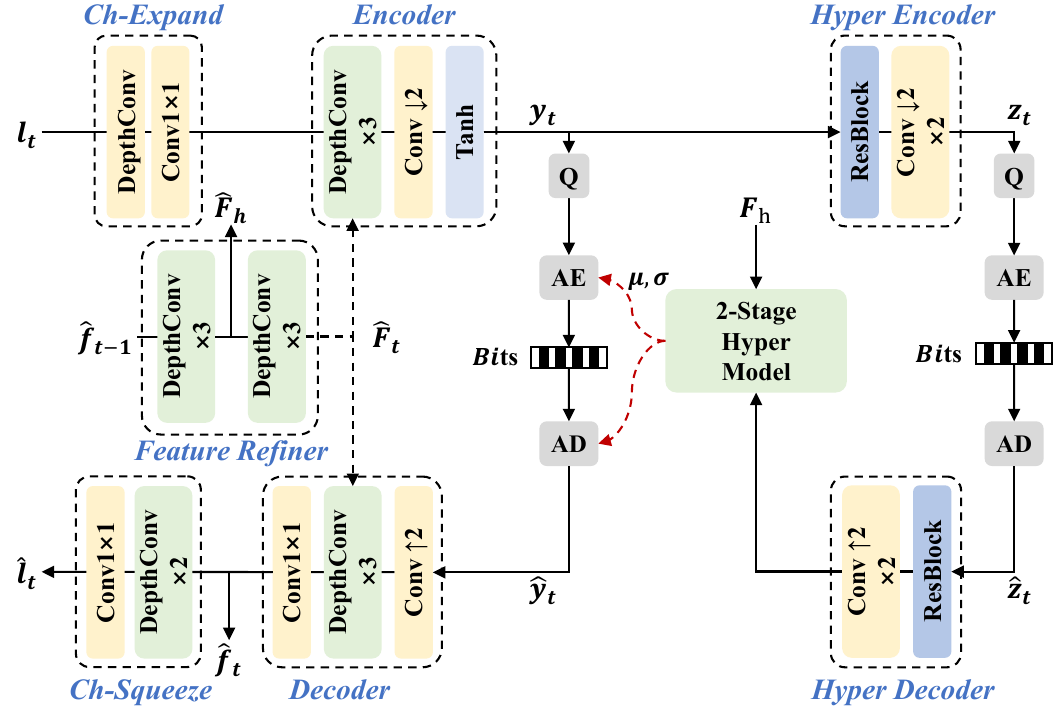}
\caption{{\bf Conditional Latent Coder (CLC)}. The previous frame’s feature \(\hat{f}_{t-1}\) is processed by depth-wise convolution blocks to produce temporal conditions \(\hat{F}_{t}\) and \(\hat{F}_h\) for the main encoder–decoder and entropy model. }
\label{fig:codec}
\end{figure}

{To further exploit the statistical redundancy in $l_t$, we apply a conditional latent coder (CLC) that follows the design of popular VAE-based conditional video coders—specifically DCVC-RT in our setup (Fig. \ref{fig:codec}). This eliminates the need for explicit motion processing by modeling probabilities directly in feature space, enabling a lightweight implementation.}

{Given the limited channel dimensionality of $l_t$, \textit{e.g.}, $d_l = 32$, we employ channel expansion (Channel-expand) in the Main Encoder and the corresponding channel squeezing (Channel-squeeze) in the Main Decoder. 
\begin{itemize}
    \item In CLC’s Main Encoder, $l_t$ is first projected to $f_t$, preserving spatial resolution while expanding the channel dimension from 32 to 256. Stacked convolutional layers then transform $f_t$ into $y_t$ for entropy coding\footnote{For entropy coding, we adopt the same two-stage model as in \cite{DCVCRT, DCVC-HEM}. }. 
    \item Correspondingly, the Main Decoder reconstructs $\hat{y}_t$, which is then mapped back to a 32-channel $\tilde{l}_t$, ready for the subsequent DiT-based denoising stage. 
\end{itemize}}

{Following the design of conditional coding, temporal references are exploited in both the main and hyper coders to capture feature-space correlations and model context. Concretely, we cache the 256-channel reconstructed feature \(\hat{f}_{t-1}\) and derive temporal conditions $\hat{F}_{t}$ and $\hat{F}_h$ using the Feature Refiner. These are used for contextual embedding in the main encoder–decoder and for entropy coding. This design enables a more compact and expressive characterization of \(l_t\) in feature space \cite{fvc}.}

\subsection{One-Step Denoising with Linear DiT}
\label{sec:one_step_dit}
To achieve high-fidelity reconstruction with minimal latency, we adopt the efficient one-step denoising strategy introduced in SANA-Sprint~\cite{sanasprint, sCM}. Given the compressed latent \(\tilde{l}\) produced by the CLC\footnote{We omit the subscript $t$ in $\tilde{l}_t$ for simplicity.}, we interpret it as a noisy state at a specific noise level and restore it in a single deterministic denoising step.

As implemented in our pipeline, the denoising process consists of three sequential steps.

First, we apply a {\it timestep mapping}, where the standard diffusion timestep \(t\) is converted to the consistency model timestep \(t_{\mathrm{scm}}\) to align the signal-to-noise ratio:
\[
t_{\mathrm{scm}} = \frac{\sin t}{\cos t + \sin t}.
\]

Next, we perform {\it velocity calibration}. The noisy latent \(\tilde{l}\) is first rescaled to obtain a preconditioned latent \(\bar{l}\), which is then fed into the DiT to produce the raw model output \(v_\theta\). This output is further transformed into the calibrated consistency velocity \(\hat{F}_\theta\) via
\[
\hat{F}_\theta = \frac{(1 - 2 t_{\mathrm{scm}})\bar{l} + (1 - 2 t_{\mathrm{scm}} + 2 t_{\mathrm{scm}}^2) v_\theta}{\sqrt{t_{\mathrm{scm}}^2 + (1 - t_{\mathrm{scm}})^2}},
\]
where \(\bar{l}\) denotes the scaled (preconditioned) latent input to the DiT and \(v_\theta\) is the raw output of the denoising network.

Finally, the scheduler uses the calibrated consistency velocity \(\hat{F}_\theta\) to project the noisy latent \(\tilde{l}\) directly onto the clean data manifold, yielding the denoised latent \(\hat{l}\) in a single consistency update.

\section{Multi-Stage Training of YODA} \label{sec:training}

To ensure training stability and performance, YODA is trained in three main steps.

\subsection{Stage I: Pretraining Temporal-Aware AutoEncoder (TA-AE)}

The proposed TA-AE is first trained with a composite distortion objective that blends pixel-wise, perceptual, and structural losses:
\begin{align}
\mathcal{D}_{\text{rec}}
= \lambda_1\,\mathcal{L}_{\text{MSE}}
+ \lambda_2\,\mathcal{L}_{\text{LPIPS}}
+ \lambda_3\,\mathcal{L}_{\text{DISTS}}.
\end{align}
$\mathcal{L}_{\rm MSE}$ enforces pixel-level fidelity via mean squared error; $\mathcal{L}_{\rm LPIPS}$  captures perceptual similarity in deep feature space using pretrained networks \cite{lpips}; and $\mathcal{L}_{\rm DISTS}$ preserves structural and textural consistency \cite{dists}. The weights $\lambda_1$, $\lambda_2$, and $\lambda_3$  balance the contributions of each term. 

The training objective of this Stage I further adds an adversarial term to promote photo-realistic reconstructions: 
\begin{align}
\mathcal{L}_{\text{Stage I}}
= \mathcal{D}_{\text{rec}}
+ \lambda_{\text{adv}}\,\mathcal{L}_{\text{adv}},
\end{align}
where $\mathcal{L}_{adv}$ is the adversarial loss and $\lambda_{adv}$ is its weight.

 Unlike conventional VAEs, our framework omits explicit KL regularization and its associated stochastic prior. Owing to the large downsampling factor (\textit{e.g.}, 32×), the resulting latents naturally follow a smooth, approximately Gaussian distribution. This, in turn, enables stable end-to-end training using only the reconstruction objective.

\subsection{Stage II: Jointly Training of Conditional Latent Coder and DiT}

After establishing a stable TA-AE, we then jointly optimize the Conditional Latent Coder (CLC) and the linear DiT-based denoiser. To preserve the DiT’s generative priors while adapting it to our specific latent manifold, we apply LoRA-based fine-tuning. The optimization objective for this stage is given by
\begin{align}
\mathcal{L}_{\mathrm{Stage\,II}}
= 
\mathcal{D}_{\mathrm{rec}}
+ \lambda_{\mathrm{rate}}\, \mathcal{R}.
\label{eq:stage2}
\end{align}

{Although Stage II effectively balances reconstruction and generative behavior, the TA-AE module remains frozen during this phase.}

\subsection{Stage III: End-to-End Fine-tuning}

By jointly training the CLC and the TA-AE under bitrate constraints, we explicitly push the autoencoder to operate effectively in low-bitrate regimes. Since lower bitrates restrict the information that can be encoded from the current frame, the TA-AE is encouraged to guide the production of more compact latent representations by more aggressively leveraging temporal correlations. This, in turn, reduces the entropy burden on the CLC while preserving reconstruction quality.

In this final phase, all components are fine-tuned end-to-end. The bitrate regularization term \(\mathcal{R}\) maintains a balance between fidelity and compression efficiency, while an adversarial loss \(\mathcal{L}_{\mathrm{adv}}\), driven by a PatchGAN~\cite{patchgan} discriminator, improves perceptual realism and alleviates the over-smoothing artifacts typical of purely pixel-wise objectives. The overall training objective is
\begin{align}
\mathcal{L}_{\mathrm{Stage\,III}}
= 
\mathcal{D}_{\mathrm{rec}}
+ \lambda_{\mathrm{rate}}\, \mathcal{R}
+ \lambda_{\mathrm{adv}}\, \mathcal{L}_{\mathrm{adv}}.
\label{eq:stage3}
\end{align}

This joint optimization harmonizes rate control, perceptual quality, and adversarial realism, yielding a unified representation space that supports both efficient encoding and high-fidelity generation.

\begin{figure}[t]
    \centering
    \includegraphics[width=0.9\linewidth]{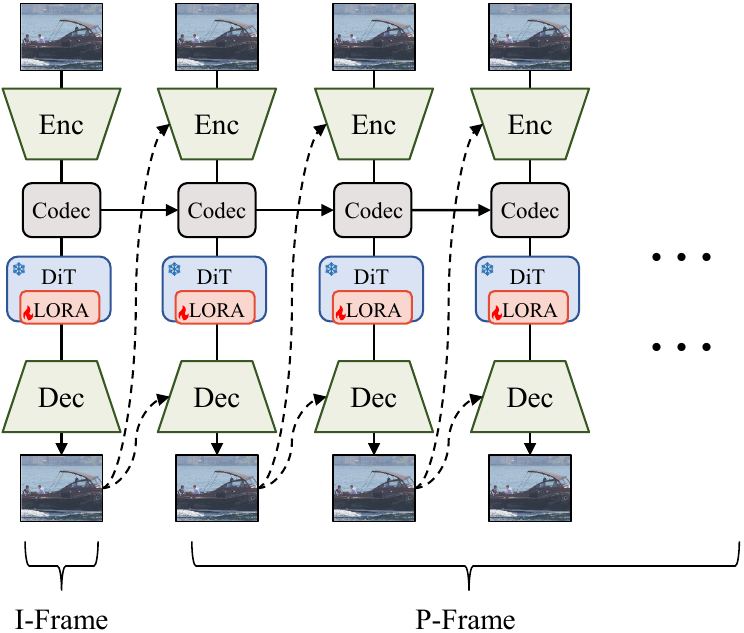}
\caption{{\bf Low-Delay IPPP Structure} used in YODA. Currently, I-Frame and P-Frame share a similar architecture and the same training.}
    \label{fig:ipp}
\end{figure}

{\bf Remarks.} {Note that the discussion above assumes the availability of temporal references. In our current setting, this corresponds to P-frames with forward prediction; B-frames with both forward and backward prediction are left for future work. By contrast, at random access points, the corresponding frames are typically encoded as I-frames, for which no temporal reference is available. In this case, we replace our TA-AE with DC-AE~\cite{dcae} and remove the temporal conditions (i.e., $\hat{F}_t$ and $\hat{F}_h$ in Fig. \ref{fig:framework}) from the CLC. A similar three-stage training strategy is applied in this I-frame setting. }

{Figure~\ref{fig:ipp} illustrates a popular IPPP structure widely used in low-delay scenarios, where the first frame at each random access point is encoded as an I-frame, and the subsequent frames are encoded as P-frames until the next random access point is reached. Such a random-access I-frame can be manually configured or content-adaptive, using a scene-detection algorithm.}

%% file: sec/4_exp.tex
\section{Experimental Results and Analysis}

\subsection{Implementation}
\textbf{Datasets.} 
We train all three stages using the Vimeo-90k dataset~\cite{vimeo90k}, utilizing 7-frame septuplets randomly cropped to a resolution of $256\times256$. 
For evaluation, we employ the UVG~\cite{UVG}, MCL-JCV~\cite{MCLJCV}, and HEVC Class B~\cite{HEVC} datasets, testing on the first 96 frames of each sequence at 1080p resolution. 
To ensure consistent color representation, all YUV input frames are converted to RGB following the ITU-R BT.709 standard prior to inference.

\begin{table*}[t]
\centering
\caption{BD-Rate (\%) comparison of different video compression methods on three datasets. H.266/VVC's reference software VTM-23.13 is used as the anchor. (\(\downarrow\)) indicates the lower the better; ``N/A'' denotes data is unavailable in the original publication.}
\label{tab:perception_distortion}
\small
\setlength{\tabcolsep}{4pt}
\renewcommand{\arraystretch}{1.1}

\begin{tabularx}{\linewidth}{ll*{3}{Y}Y}
\toprule
\multirow{2}{*}{Dataset} & \multirow{2}{*}{Methods} 
& \multicolumn{4}{c}{Metrics} \\
\cmidrule(lr){3-6}
& & DISTS $\downarrow$ & LPIPS $\downarrow$ & KID $\downarrow$ & FID $\downarrow$ \\
\midrule
\multirow{7}{*}{UVG} 
& HM-18.0    & +10.94 & +54.82 & +104.51 & +36.48 \\ 
& DCVC-RT    & +0.62  & -21.05 & +4.53 & +23.91 \\ 
& PLVC       & -79.31 & -89.87 & -89.55 & -19.36 \\ 
& GLC-video  & -90.74 & -95.38 & N/A & N/A \\ 
& DiffVC     & -88.29 & -81.71 & -72.41 & N/A \\        
\rowcolor{OursBlue}
\cellcolor{white}
& \textbf{Ours} &  \textbf{-98.60}&  \textbf{-96.83} & \textbf{-99.30} &  \textbf{-96.49}  \\
\midrule
\multirow{7}{*}{HEVC-B} 
&  HM-18.0    & +5.05  & +51.48
& +60.94    & +24.50  \\ 
& DCVC-RT     & +8.18 &  +31.37
& +41.40 &  +29.25  \\ 
& PLVC        & -78.92 & -82.38
& -12.06 & -3.18  \\ 
& GLC-video   &-86.92 & -91.94  & N/A & N/A  \\
        \rowcolor{OursBlue}
\cellcolor{white}
& \textbf{Ours}        &  \textbf{-98.24} &   \textbf{-95.67}  &   \textbf{-98.25}  &  \textbf{-94.34}  \\
\midrule
\multirow{7}{*}{MCL-JCV} 
&  HM-18.0   & +15.26  & +53.79 & +148.91  & +80.34\\ 
& DCVC-RT     & +11.12 & -8.39 & -23.10 & -1.07 \\ 
& PLVC        & -38.72 &-61.31&  -52.28 &  -1.54\\ 
& GLC-video   & -86.25  & -91.61 & N/A& N/A \\ 
& DiffVC      & -71.80 & -73.40 &  -18.78 & N/A \\
& DiffVC-OSD      & -83.46  & -84.38 & N/A & -35.51 \\
        \rowcolor{OursBlue}
        \cellcolor{white}
& \textbf{Ours}        & \textbf{-94.70} & \textbf{-93.92} & \textbf{-95.24} & \textbf{-94.33} \\
\bottomrule
\end{tabularx}
\label{bd-lpips}
\end{table*}

\textbf{Training Details.} 
In the first stage, we train the Temporal-Aware Autoencoder (TA-AE) by setting the reconstruction loss weights to $\lambda_1 = \lambda_2 = \lambda_3 = 1.0$ and the adversarial loss weight to $\lambda_{\mathrm{adv}} = 0.1$. 
Crucially, these hyperparameter values remain constant across all three training stages.
We train four independent models to verify performance across multiple bitrates, with rate-control weights $\lambda_{\text{rate}} \in \{1.0,\, 2.0,\, 4.0,\, 8.0\}$ ranging from high to low.
During Stage~II, we implement a progressive training strategy by gradually increasing the temporal window from 2 to 7 frames, with the learning rate initialized at $1\times10^{-4}$ and subsequently decayed to $5\times10^{-6}$. 
Finally, in Stage~III, we perform global end-to-end fine-tuning with the learning rate fixed at $5\times10^{-6}$, integrating the adversarial training objective using the constant $\lambda_{\mathrm{adv}}$ defined above.

\begin{figure*}[htbp]
\centering
\newcommand{\colwidth}{0.31} 
\includegraphics[width=0.8\linewidth]{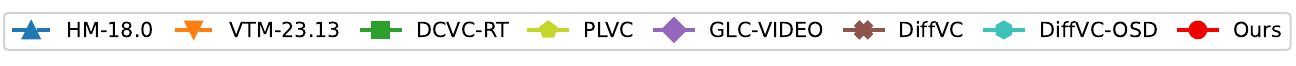} \\
\includegraphics[width=\colwidth\linewidth]{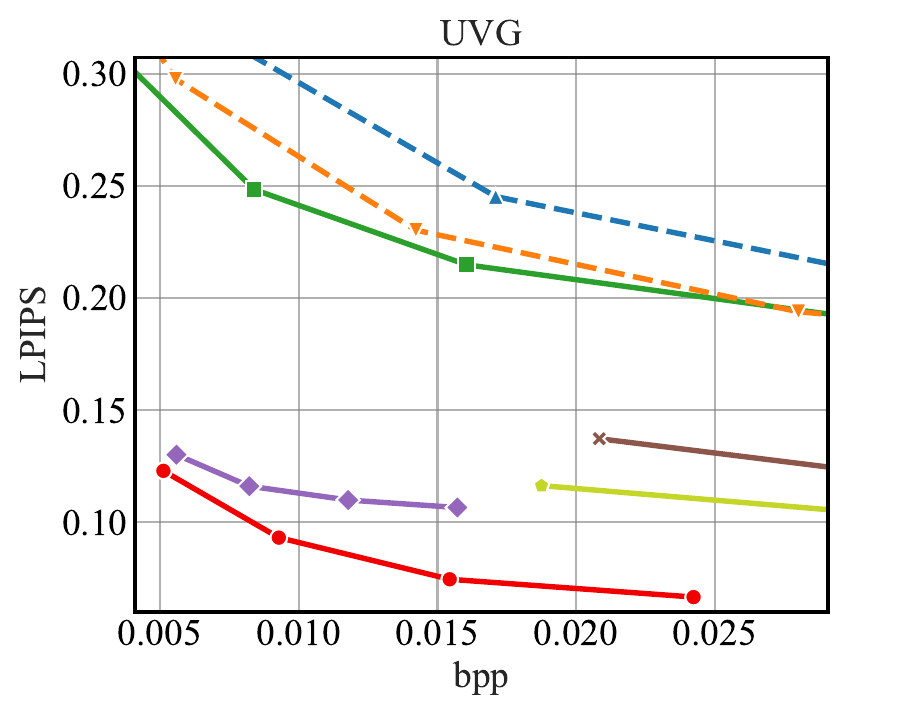} \vspace{-1mm}
\includegraphics[width=\colwidth\linewidth]{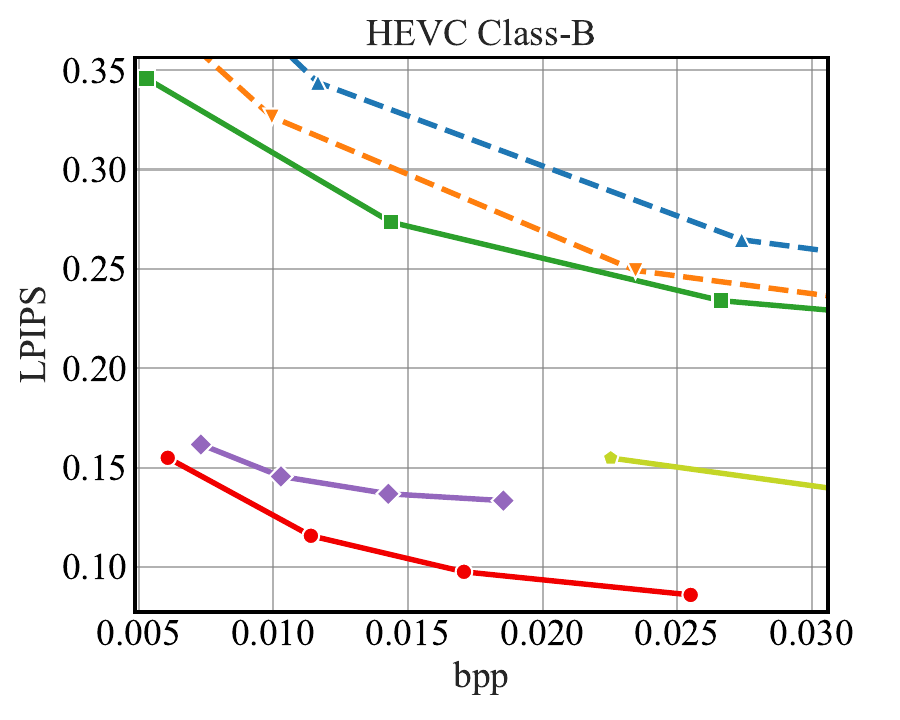} \vspace{-1mm}
\includegraphics[width=\colwidth\linewidth]{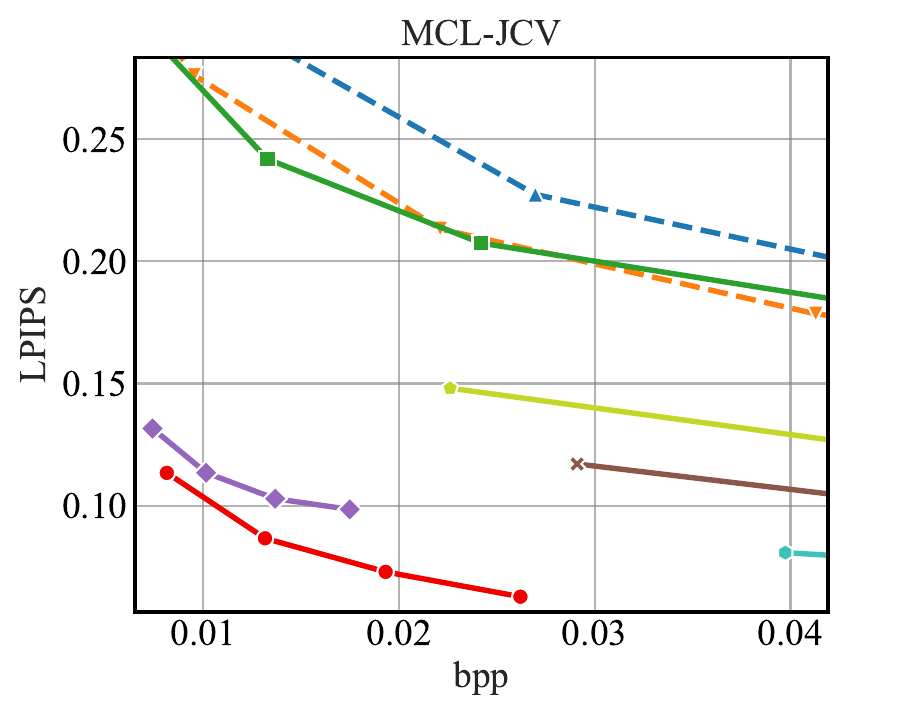} \\[1mm]
\includegraphics[width=\colwidth\linewidth]{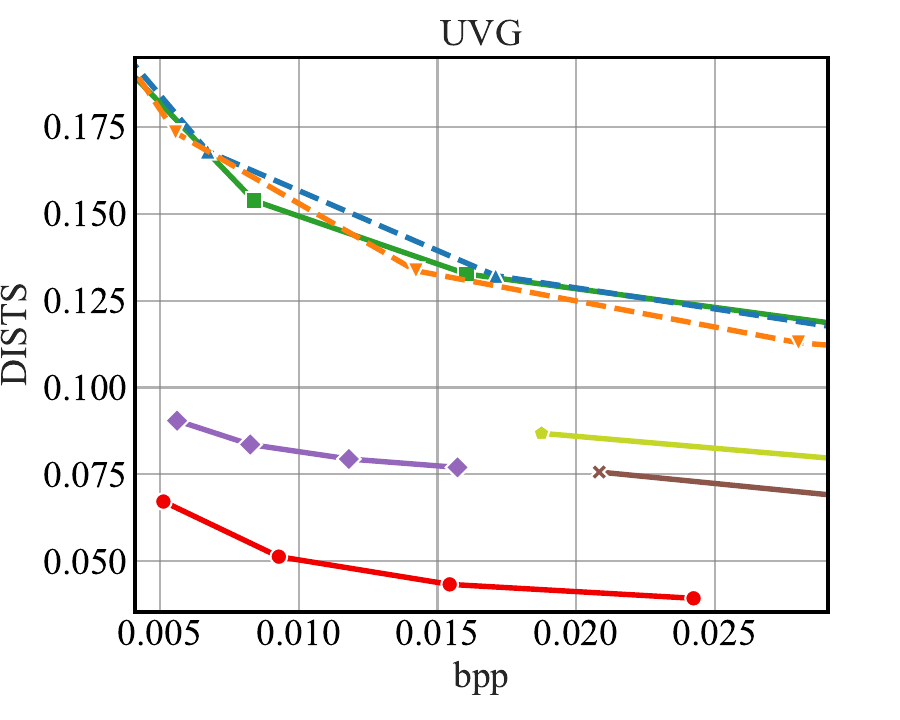} \vspace{-1mm}
\includegraphics[width=\colwidth\linewidth]{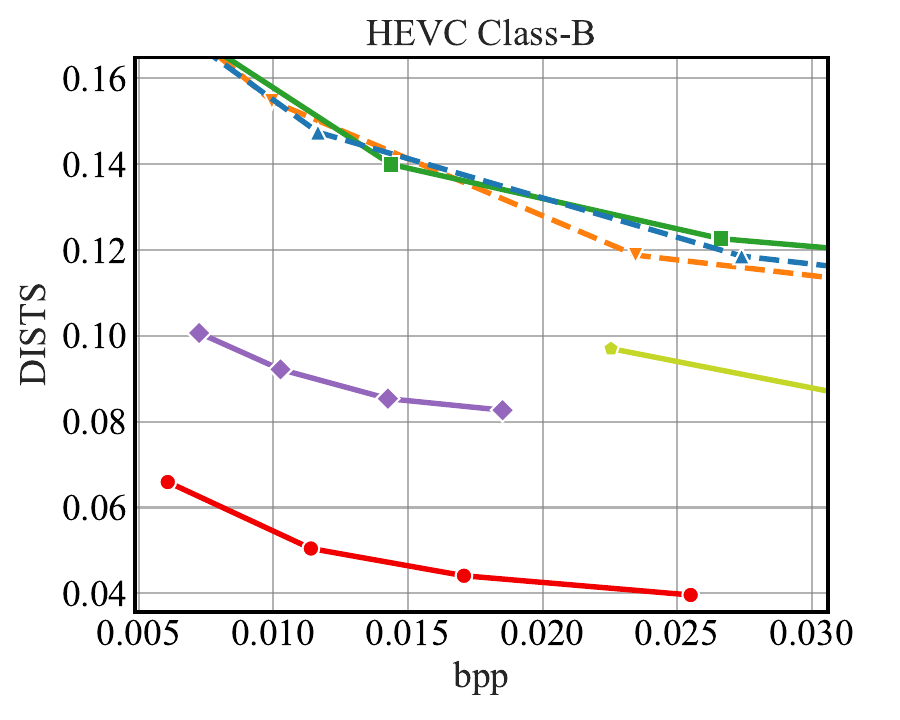} \vspace{-1mm}
\includegraphics[width=\colwidth\linewidth]{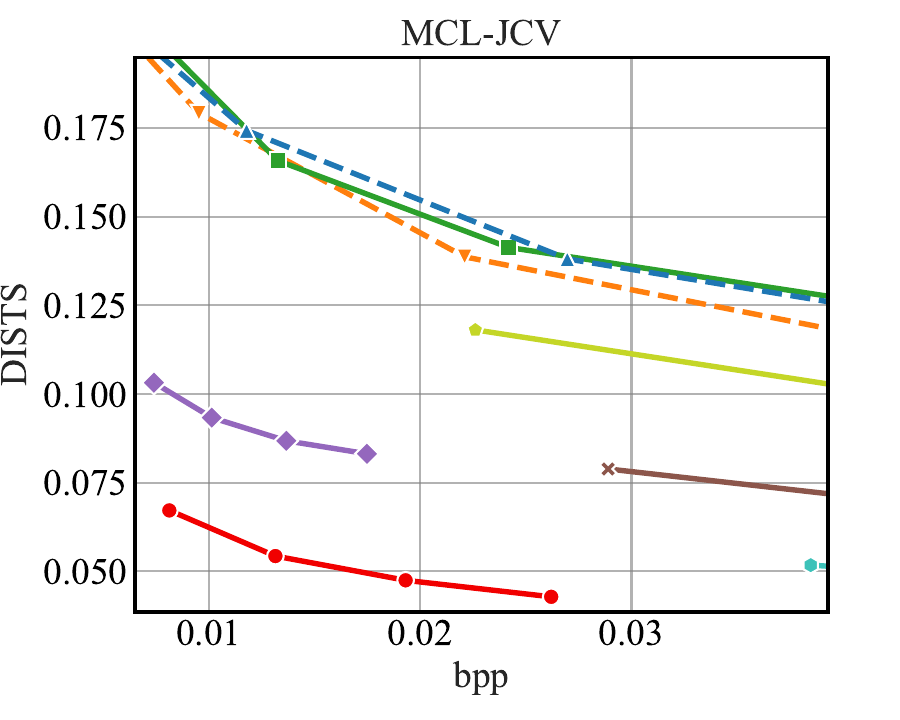}\\[1mm]
\includegraphics[width=\colwidth\linewidth]{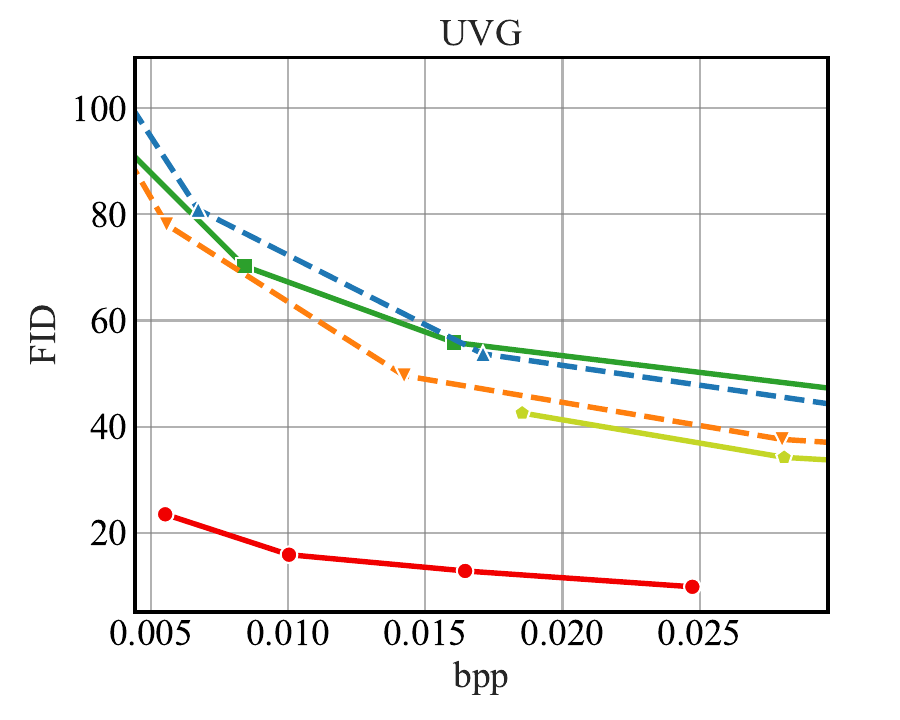} \vspace{-1mm}
\includegraphics[width=\colwidth\linewidth]{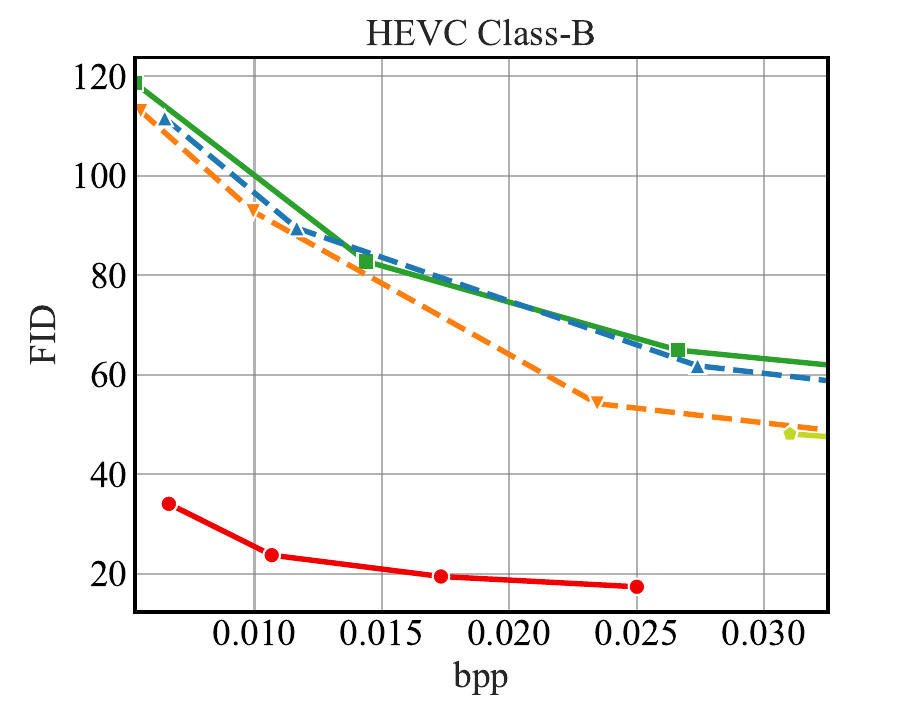} \vspace{-1mm}
\includegraphics[width=\colwidth\linewidth]{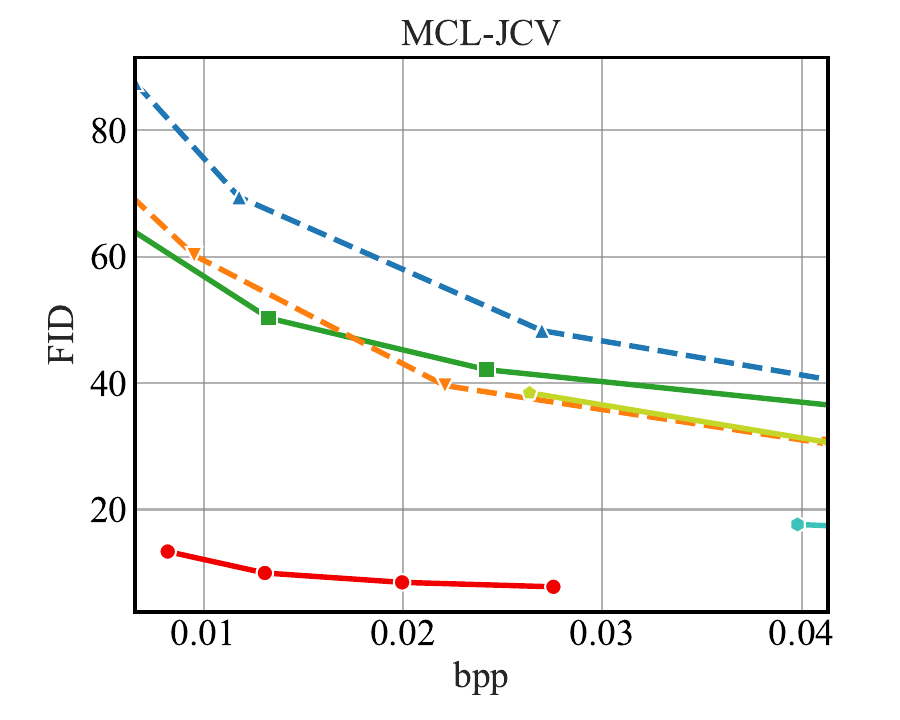}\\[1mm]
\includegraphics[width=\colwidth\linewidth]{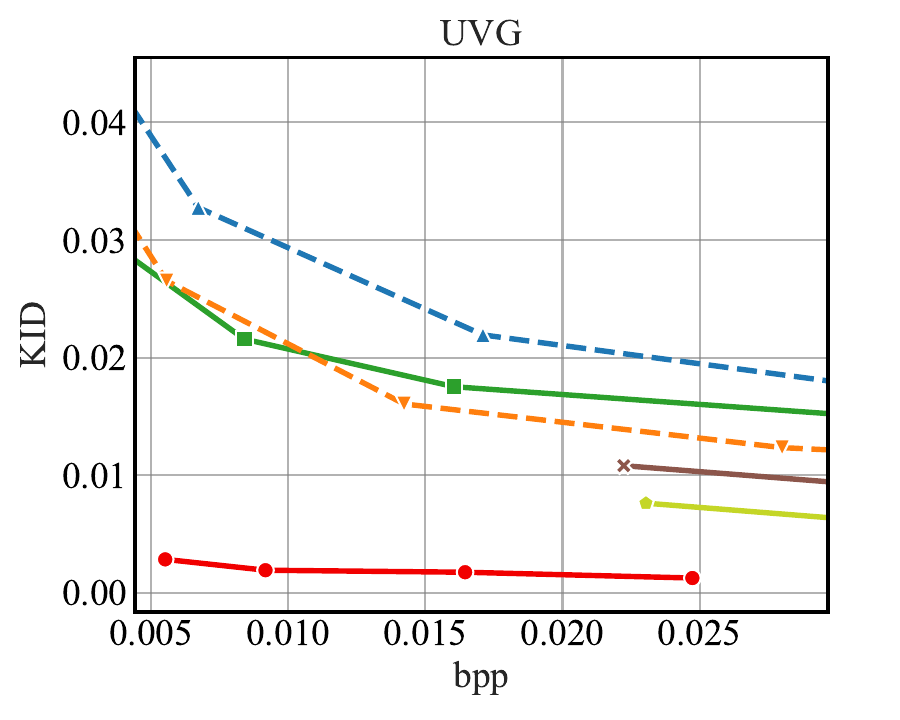} \vspace{-1mm}
\includegraphics[width=\colwidth\linewidth]{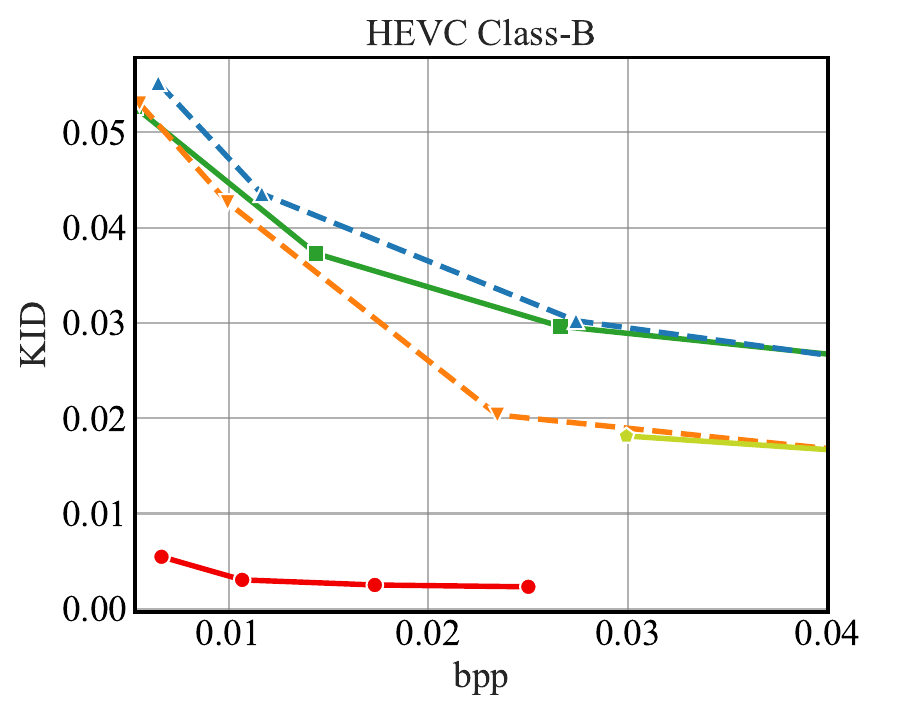} \vspace{-1mm}
\includegraphics[width=\colwidth\linewidth]{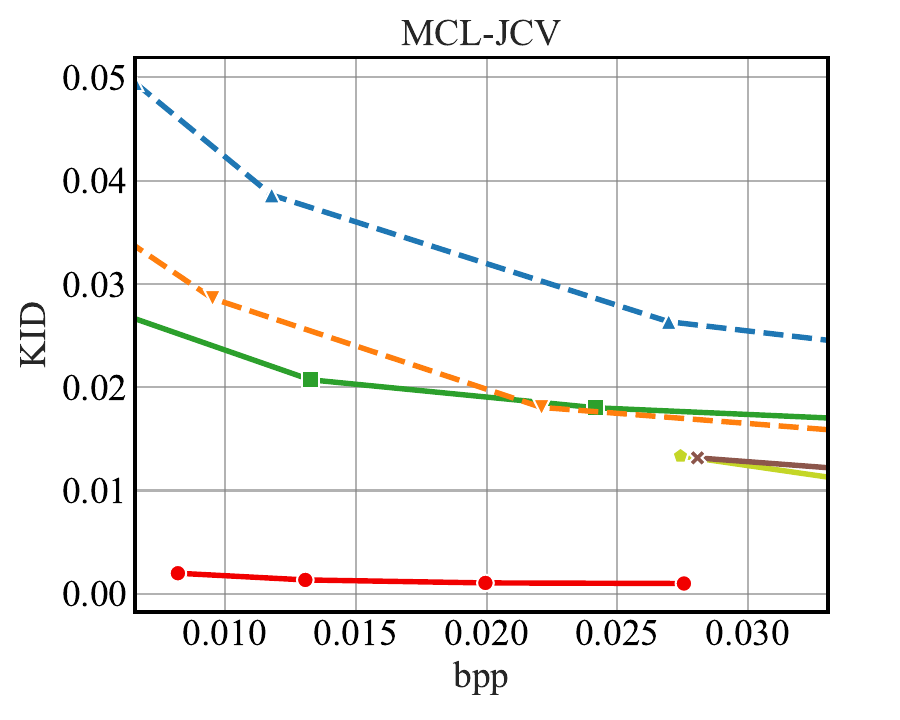}
\vspace{1mm}
\caption{Perceptual quality performance (LPIPS, DISTS, FID, and KID) comparison with other methods on UVG, HEVC-B, and MCL-JCV datasets.  The lower metric indicates better performance.}
\label{fig:compare_perceptual}
\end{figure*}

\begin{figure}[t]
\centering
\newcommand{\colwidth}{0.49} 

\includegraphics[width=0.88\linewidth]{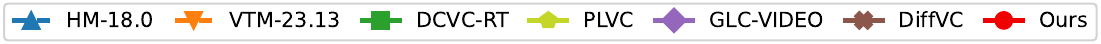} \\
\includegraphics[width=\colwidth\linewidth]{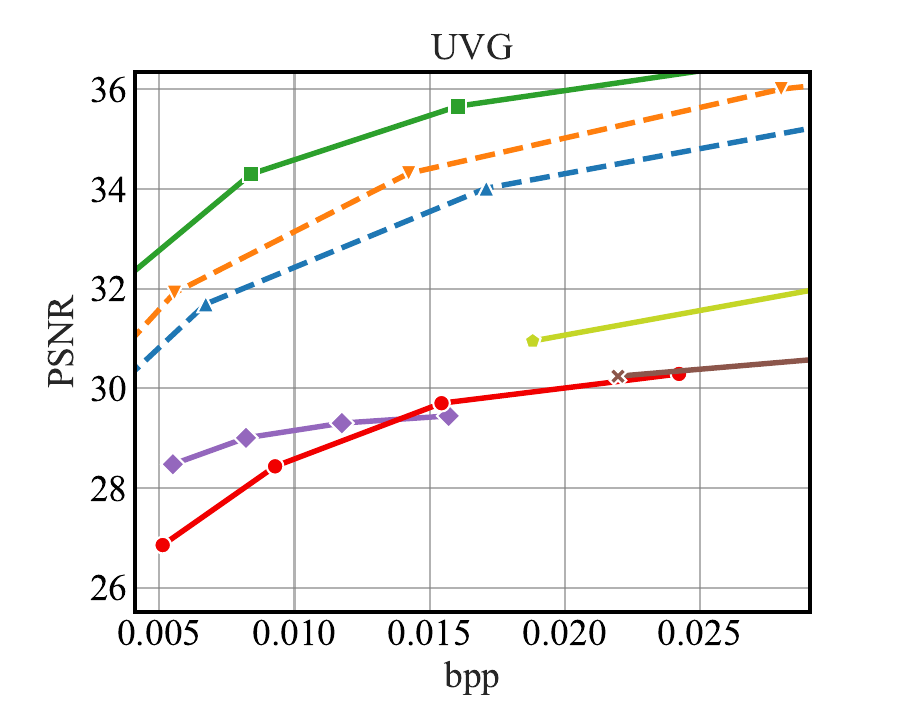} 
\includegraphics[width=\colwidth\linewidth]{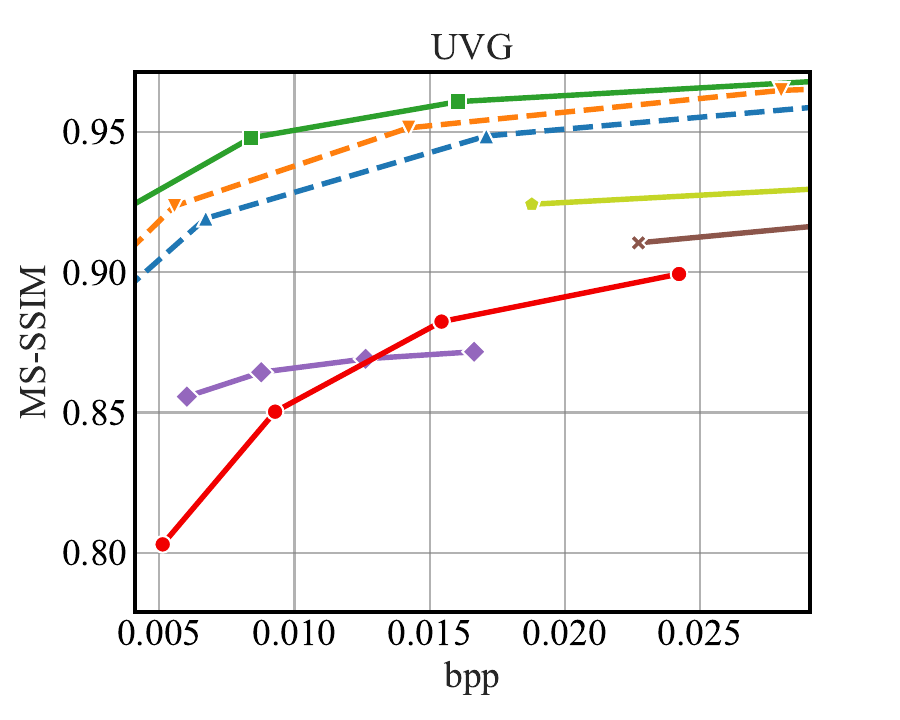}\\

\includegraphics[width=\colwidth\linewidth]{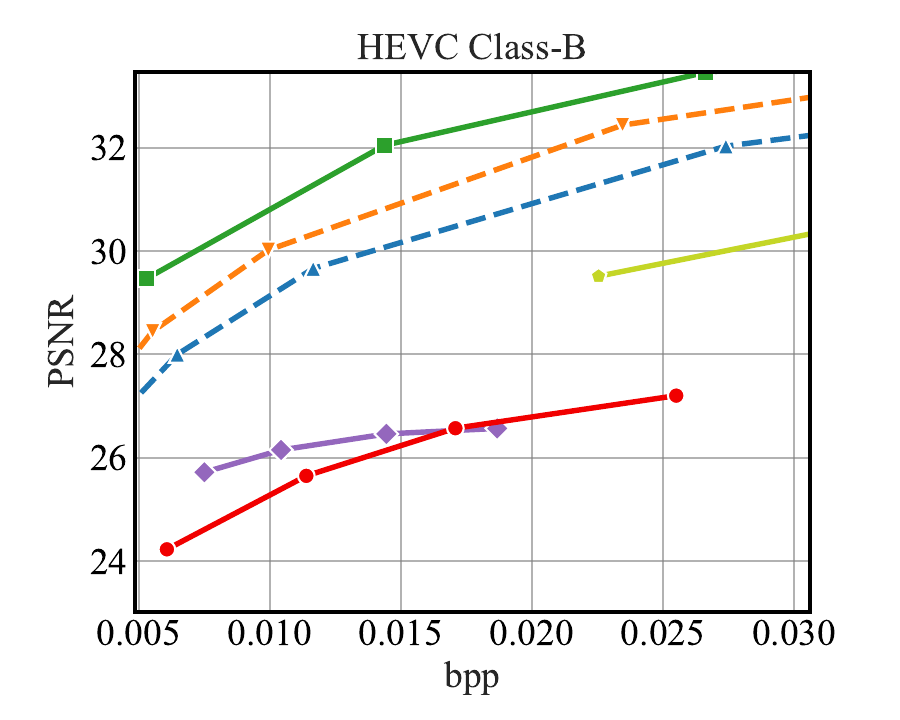}
\includegraphics[width=\colwidth\linewidth]{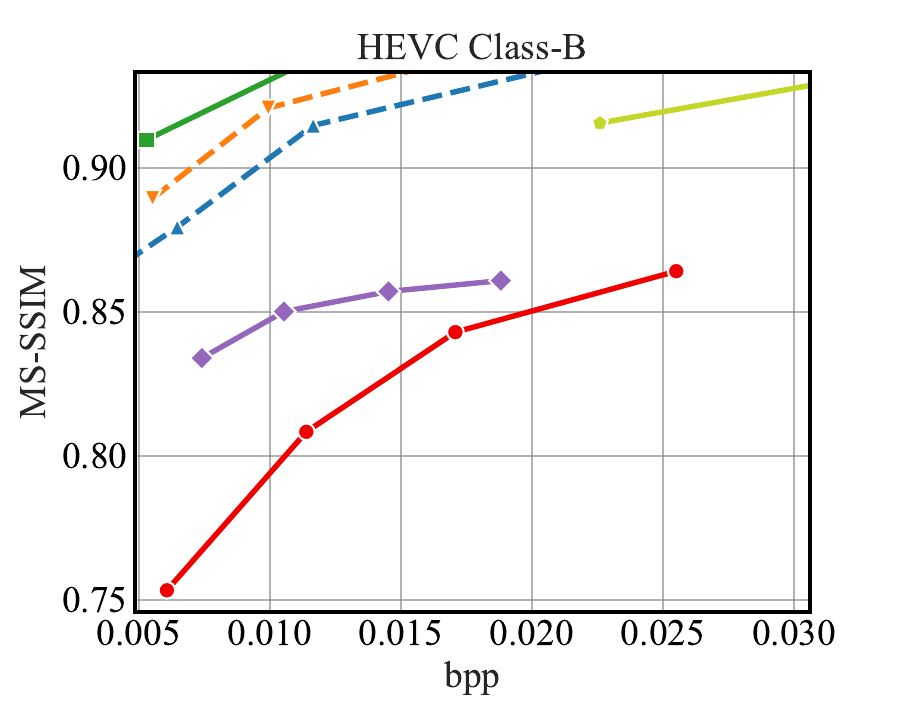}\\

\includegraphics[width=\colwidth\linewidth]{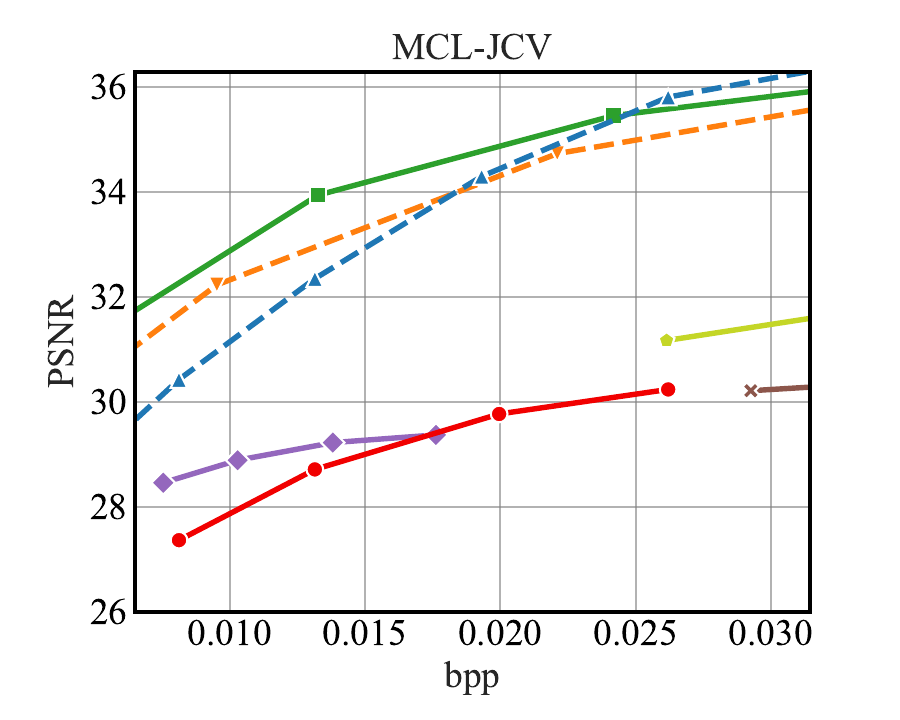}
\includegraphics[width=\colwidth\linewidth]{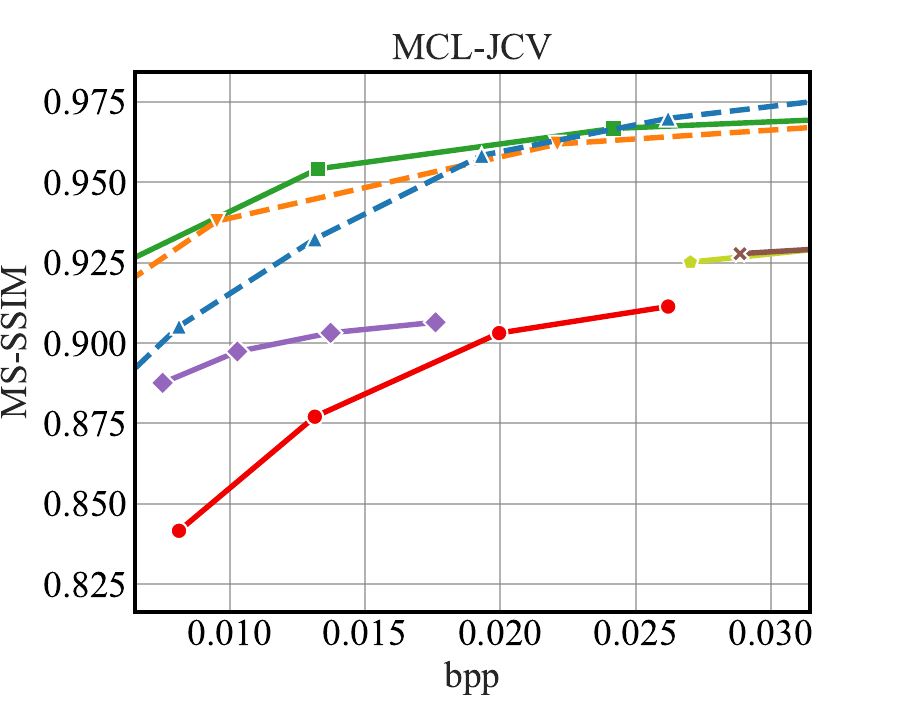}\\
\caption{Objective performance (PSNR and MS-SSIM) comparison with other methods on UVG, HEVC-B, and MCL-JCV datasets.}
\label{fig:compare_psnr_msssim}
\end{figure}

{\bf Compared Methods.} {We benchmark against traditional codecs (H.265/HEVC's reference software HM 18.0~\cite{HM}, H.266/VVC's reference software VTM 23.13~\cite{VTM}), MSE-optimized methods (DCVC-RT~\cite{DCVCRT}), GAN-based perceptual methods  (PLVC~\cite{PLVC}, GLC-VIDEO~\cite{GLCvideo}), and recent diffusion-based perceptual approaches (DiffVC~\cite{DiffVC}, DiffVC-OSD~\cite{Diffvcosd}). 
For baselines without released code, such as GLC-VIDEO, DiffVC, and DiffVC-OSD, we report the numerical results cited from their papers to retain their best performance. Visual comparisons are included for open-source frameworks, specifically DCVC-RT and VTM.}

{\bf Metrics.} {For perceptual quality (realism), we use LPIPS (Learned Perceptual Image Patch Similarity)~\cite{lpips} and DISTS (Deep Image Structure and Texture Similarity)~\cite{dists}. To quantify distributional differences between reconstructions and ground truth, we compute FID (Fr\'echet Inception Distance)~\cite{fid} and KID (Kernel Inception Distance)~\cite{kid}. We also report objective distortion metrics: PSNR and MS-SSIM. BD-Rate is used for each distortion metric as a quantitative measure for compression performance evaluation \cite{barman2022revisiting}.}

To ensure consistent evaluation, we standardize the YUV-to-RGB conversion process, as different colorimetric standards (\textit{e.g.}, BT.709 vs. BT.601) can affect metric results. Specifically, the ITU-R BT.709 standard is applied to both ground-truth and reconstructed sequences for all methods before metric computation. This ensures that performance comparisons are conducted under a unified color space definition.

\begin{figure*}[t]
    \centering
    \includegraphics[width=1\linewidth]{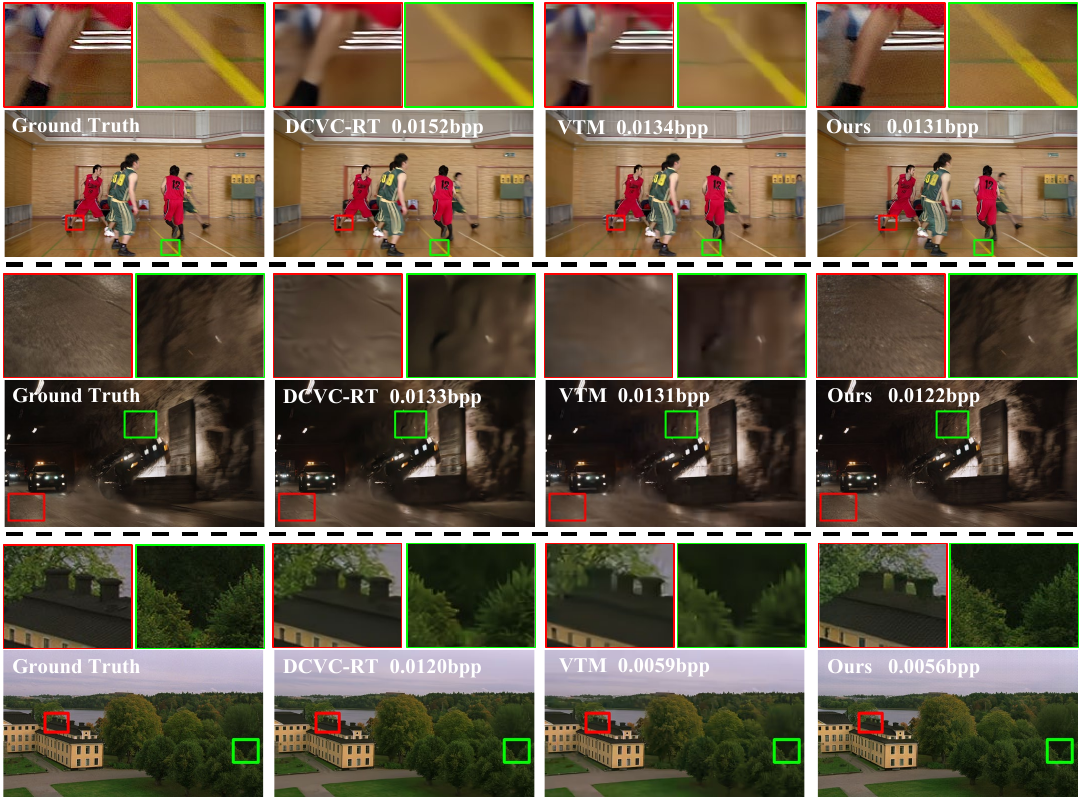}
\caption{Visual quality comparison with other methods, demonstrating the effectiveness of our approach with a lower bpp.}
    \label{fig:Qualitative}
\end{figure*}

\subsection{Results}
\textbf{Quantitative comparisons} are primarily presented to evaluate perceptual quality using LPIPS, DISTS, FID, and KID (Fig.~\ref{fig:compare_perceptual}), since diffusion-based video codecs are designed for rate-constrained perceptual realism. Nevertheless, we also report objective fidelity metrics (PSNR and MS-SSIM) in Fig.~\ref{fig:compare_psnr_msssim} to complement the main perceptual results.

As shown in Fig.~\ref{fig:compare_perceptual}, our method consistently outperforms existing approaches across all three datasets, achieving the best scores on all evaluated perceptual metrics. These results demonstrate the superiority of our model in preserving perceptual similarity and texture fidelity (LPIPS/DISTS), while also ensuring distributional consistency and visual realism (FID/KID).

In Table~\ref{bd-lpips}, we report BD-rate gains for the perceptual metrics DISTS, LPIPS, KID, and FID, using VTM-23.13 as the anchor. The tabulated results show that our method substantially outperforms the baselines in perceptual quality, consistently achieving lower BD-Rate values across all these indicators.

\textbf{Qualitative Comparisons.}
As shown in Fig.~\ref{fig:Qualitative}, We compare our method with DCVC-RT, a representative recent neural video codec, and the traditional VTM-23.13 codec. At low bitrates, both baselines exhibit pronounced blurring artifacts. In contrast, even at lower bitrates, our model better preserves the structure of dynamic content (\textit{e.g.}, the athlete’s leg) and maintains background textures such as ground details, yielding reconstructions that are visually closer to the ground truth.

\textbf{Complexity Analysis.}
{Table~\ref{tab:complexity} further summarizes the complexity of YODA using model size and decoding latency for I-Frame and P-Frame, respectively. For model size, we present the specific parameters of each modular component and the trainable parts.}

{As shown, P-Frame decoding is approximately \(1.5\times\) slower than I-frame decoding at the 1080p spatial resolution. This additional latency primarily stems from the TA-AE: the extra feature extraction branch used to fuse multiscale temporal conditions increases both computational overhead and parameter count. The complexity analysis suggests that an interesting avenue for future exploration is to reduce the decoding latency for real-time processing.}

\begin{table}[t]
  \centering
\caption{Complexity analysis of YODA. Decoding latency is measured on an NVIDIA 5090 GPU. ``Trainable Params'' refers to the parameters involved in the final joint training.}
  \label{tab:complexity}
  \renewcommand{\arraystretch}{1.1}
  \setlength{\tabcolsep}{3.5pt}
  \resizebox{\linewidth}{!}{
  \begin{tabular}{c|c|cc}
    \toprule
    \textbf{Category} & \textbf{Item} & \textbf{I-Frame} & \textbf{P-Frame} \\
    \midrule
    \multirow{1}{*}{\textbf{Inference Speed (s)}} & Decoding Latency (1080p) & 0.665 & 1.028 \\
    \midrule
    \multirow{3}{*}{\textbf{Modular Params (M)}} & AutoEncoder & 312.25  & 413.57 \\
    & Latent Coder & 44.13 & 18.83 \\
    & DiT Denoiser & 630.75 & 630.75 \\
    \midrule
    \multirow{2}{*}{\textbf{Model Size (M)}}
    & Total Params & 987.12 & 1063.15 \\
    & Trainable Params & 393.62 & 469.65 \\
    \bottomrule
  \end{tabular}}
\end{table}

\begin{figure}[t]
    \centering
    \subfloat[]{
        \label{fig:ablation_lpips_taae}
        \includegraphics[width=0.46\linewidth]{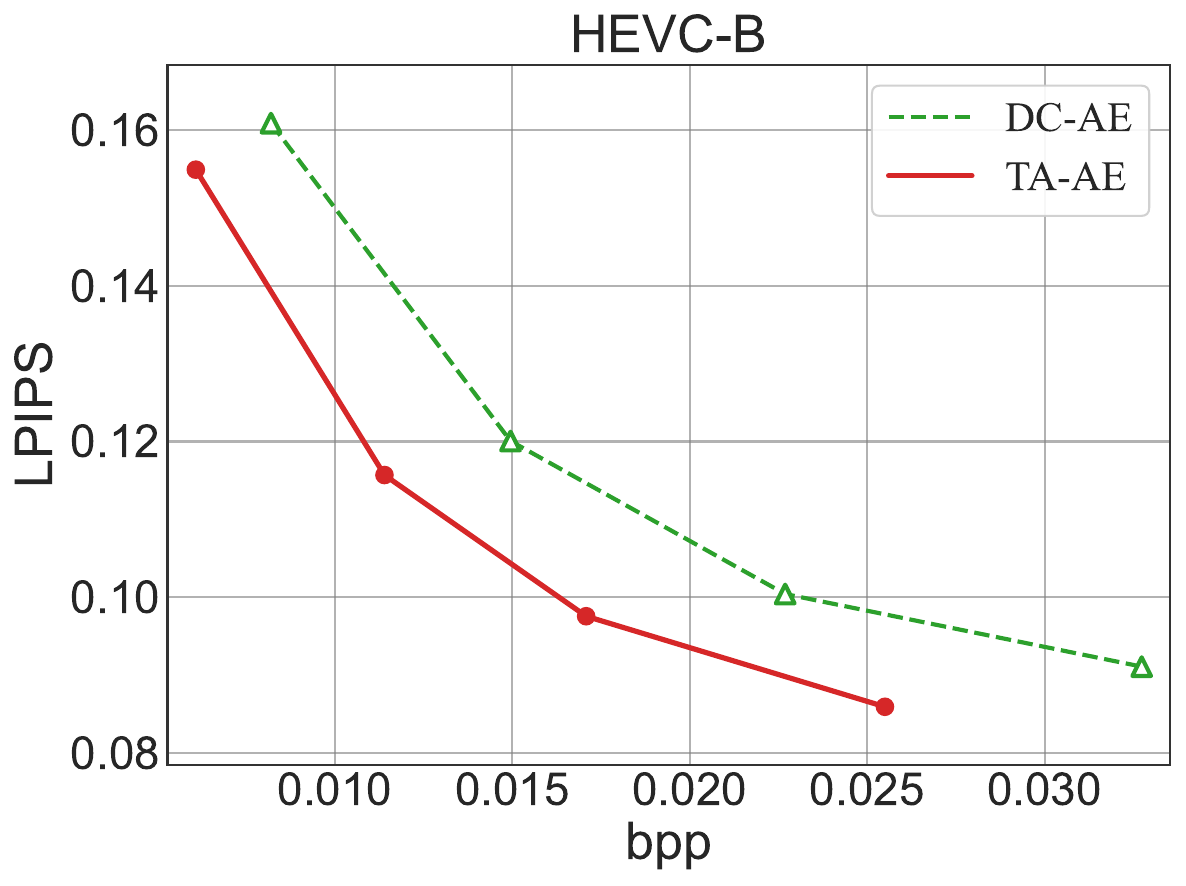}
    }
    \hfill 
    \subfloat[]{
        \label{fig:ablation_dists_taae}
        \includegraphics[width=0.46\linewidth]{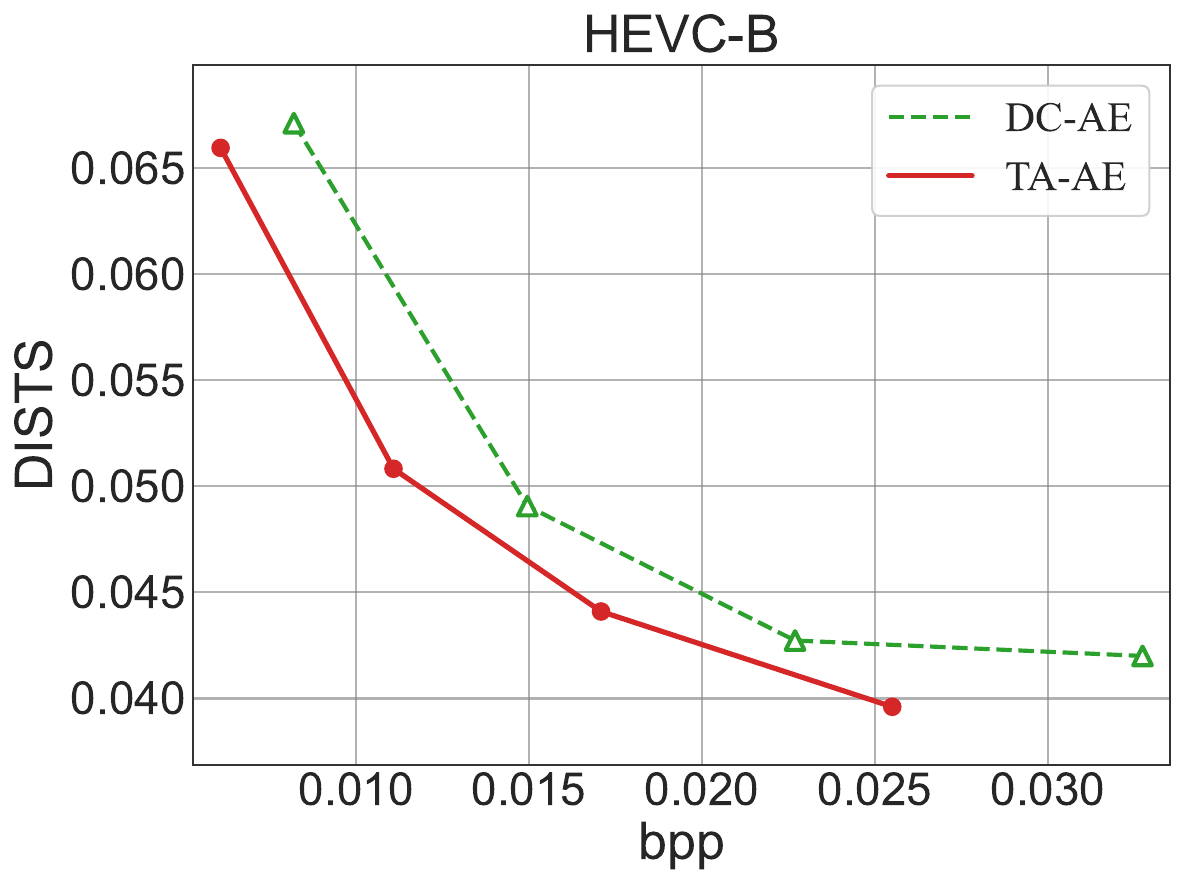}
    }
    \caption{Temporal-Aware AutoEncoder (TA-AE) versus frame-independent DC-AE for latent generation. HEVC Class B samples are used for evaluation.}
\label{fig:ta-ae-hevcb}
\end{figure}

We next conduct a thorough ablation study to analyze the contribution of each component and better understand the capability of the proposed YODA framework.

\subsection{Ablation Studies}

\subsubsection{Temporal-Aware AutoEncoder (TA-AE)}

\paragraph{Impact of temporal awareness}

To assess the contribution of temporal awareness, we remove it and revert to the default DC-AE, which performs purely spatial encoding for each frame. In this configuration, the autoencoder processes each frame independently, without access to multiscale features from the reference frame, while the remaining components (CLC and DiT) and the training are kept unchanged. Figure~\ref{fig:ta-ae-hevcb} illustrates the resulting performance gap, clearly showing that the proposed TA-AE with temporal information embedding consistently outperforms DC-AE. Corresponding BD-Rate measures can be seen in the first row of Table \ref{tab:context_scales_only} (zero scales using DC-AE).

\paragraph{Impact of multiscale embedding of temporal condition}  
By default, the reconstructed reference frame is processed and embedded across five scales in the autoencoder as temporal conditions in TA-AE (see \(\{\hat{c}_i\}_{i=1}^5\) in Fig.~\ref{fig:temporalAE}). Here, we quantitatively assess the contribution of these scales by incrementally enabling them. Specifically, we retrain models while selectively activating particular temporal scales in TA-AE, keeping the remaining architecture unchanged. Table~\ref{tab:context_scales_only} reports the resulting BD-rate values using LPIPS and DISTS as distortion measures for these configurations. We also include the corresponding decoding latency at 1080p and 480p.

As shown in Table~\ref{tab:context_scales_only}, introducing the first high-resolution scale (\(\hat{c}_1\)) yields the most significant improvement, reducing the LPIPS BD-rate from +45.58\% (DC-AE baseline without using any temporal condition) to +14.19\%. A second substantial gain is observed when enabling the first three scales (\(\hat{c}_1, \hat{c}_2, \hat{c}_3\)), with the BD-rate further decreasing to +1.74\%. Extending the temporal context from three to five scales yields diminishing returns in distortion reduction (LPIPS improves only from +1.74\% to 0.00\%). Although adding \(\hat{c}_4\) and \(\hat{c}_5\) does not provide significant BD-rate gains, the additional decoding latency remains negligible. Consequently, we adopt the five-scale configuration in this work, whereas future applications may flexibly adjust the number of temporal scales based on their latency and complexity constraints.

\begin{table}[t]
    \centering
\caption{BD-Rate and decoding latency by embedding different scales of temporal conditions. ``\checkmark'' denotes the inclusion of the corresponding scale. Anchor uses 5 scales in default.}
    \label{tab:context_scales_only}
    \setlength{\tabcolsep}{2mm}
    \resizebox{\linewidth}{!}{
    \begin{tabular}{lccccc|cc|cc}
        \toprule
        \multirow{2}{*}{\textbf{Feature Levels} ($N$)} 
            & \multicolumn{5}{c}{\textbf{Scales}} 
            & \multicolumn{2}{c}{\bf Decoding Latency} 
            & \multicolumn{2}{c}{\textbf{BD-Rate}} \\
        \cmidrule(lr){2-6} \cmidrule(lr){7-8} \cmidrule(lr){9-10}
         & $\hat{c}_1$ & $\hat{c}_2$ & $\hat{c}_3$ & $\hat{c}_4$ & $\hat{c}_5$ & 1080p & 480p & LPIPS $\downarrow$ & DISTS $\downarrow$ \\
        \midrule
        0 scale (DC-AE)  & \checkmark & & & & & 0.657s & 0.195s & +45.58\% & +26.68\% \\
        \addlinespace
        1 Scale  & \checkmark & & & & & 0.831s & 0.228s & +14.19\% & +16.7\% \\
        \addlinespace
        2 Scales & \checkmark & \checkmark & & & & 0.926s & 0.252s & +13.40\% & +13.5\% \\
        \addlinespace
        3 Scales & \checkmark & \checkmark & \checkmark & &  & 1.003s & 0.262s & +1.74\% & +1.34\% \\
        \addlinespace
        4 Scales & \checkmark & \checkmark & \checkmark & \checkmark & & 1.021s & 0.267s & +1.04\% & +0.95\% \\
        \addlinespace
        \rowcolor{OursBlue}
        5 Scales & \checkmark & \checkmark & \checkmark & \checkmark & \checkmark & 1.028s & 0.269s & 0.00\% & 0.00\% \\
        \bottomrule
    \end{tabular}}
\end{table}

\paragraph{Long-term Temporal Referencing} We generally assume one temporal reference frame in this work for generating temporal conditional priors.  Here, we investigated the inclusion of an additional temporal reference to capture longer-range dependencies and improve performance. Specifically, {we fused the information corresponding to two previous reconstructed frames, e.g., image-space \(\hat{x}_{t-1}\) and \(\hat{x}_{t-2}\) for TA-AE, and feature-space \( \hat{f}_{t-1}, \hat{f}_{t-2} \) for CLC.} However, experimental results show that incorporating an additional reference frame yields a BD-Rate change of less than 1\% relative to the single-frame baseline. This indicates that the immediate previous frame already provides the dominant temporal information. Since the multi-frame design increases feature-processing costs while yielding only marginal gains, we adopt the more straightforward single-reference strategy in this work.

\subsubsection{Conditional Latent Coder (CLC)}

\paragraph{Impact of internal channel expansion}
Recalling that we expand the channel dimensionality of the latent features produced by TA-AE from 32 to 256 (i.e., from \(l_t\) to \(f_t\)), we now examine different channel configurations to understand the contribution.

As shown in Table~\ref{tab:ablation_channels}, setting \(f_t\) with 256 channels yields a substantial quality improvement. Beyond this point, the gains quickly saturate: using 512 channels provides almost no additional benefit (–0.03\% LPIPS). At the same time, reducing the number of channels to 32 does not lead to a meaningful speedup, since the latent features are already highly downsampled (spatial resolution of \(1/32\)). Consequently, we adopt \(C = 256\) as a balanced choice that delivers high perceptual quality without introducing unnecessary complexity.


\paragraph{Embedding position of temporal features}

\begin{table}[t]
  \centering
  \caption{BD-Rate \& decoding time comparison using HEVC Class B  sample. $\Delta$Dec Time measures the relative change in decoding time compared to the anchor using 256 channels.}
  \label{tab:ablation_channels}
  \renewcommand{\arraystretch}{1.3}
  \setlength{\tabcolsep}{4pt}
  \begin{tabular}{lccccc}
    \toprule
    \multirow{2}{*}{\textbf{Metric}} & \multicolumn{5}{c}{{Channels of $f_t$}} \\
    \cmidrule(lr){2-6}
     & 32 & 64 & 128 & \textbf{256 (Ours)} & 512 \\
    \midrule
    {LPIPS} & +42.71\% & +26.36\% & +11.11\% & \textbf{0.00\%} & -0.03\% \\
    {DISTS} & +55.02\% & +27.01\% & +13.48\% & \textbf{0.00\%} & -0.02\% \\
    \midrule
    \textbf{$\Delta$Dec Time}   & $-0.10\%$  & $ -0.08\%$ & $ -0.05\%$ & \textbf{0.00\%} & +0.05\% \\
    \bottomrule
  \end{tabular}
\end{table}

\begin{figure}[htbp]
    \centering
        
    \subfloat[]{
        \includegraphics[width=0.8\linewidth]{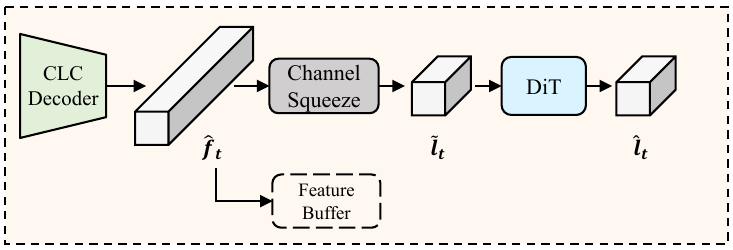} \label{fig:compare_framework2}
    }
    \hfill 
    \subfloat[]{
        \includegraphics[width=0.8\linewidth]{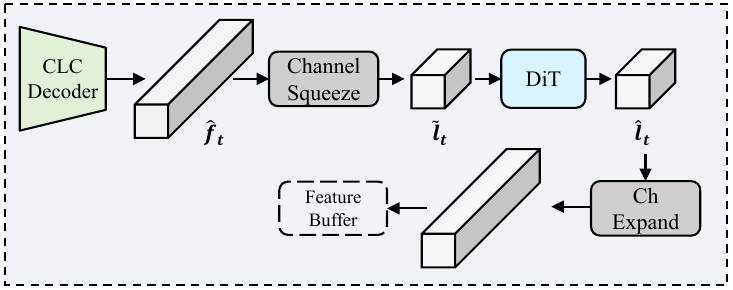}  \label{fig:compare_framework1}
    }
        \caption{Embedding position of temporal priors used in CLC: (a) Pre-DiT; (b) Post-DiT.}
    \label{fig:compare_framework_post_pre}
\end{figure}

\begin{figure}[ht]
    \centering
    \subfloat[]{
        \label{fig:ablation_lpips_prop}
        \includegraphics[width=0.47\linewidth]{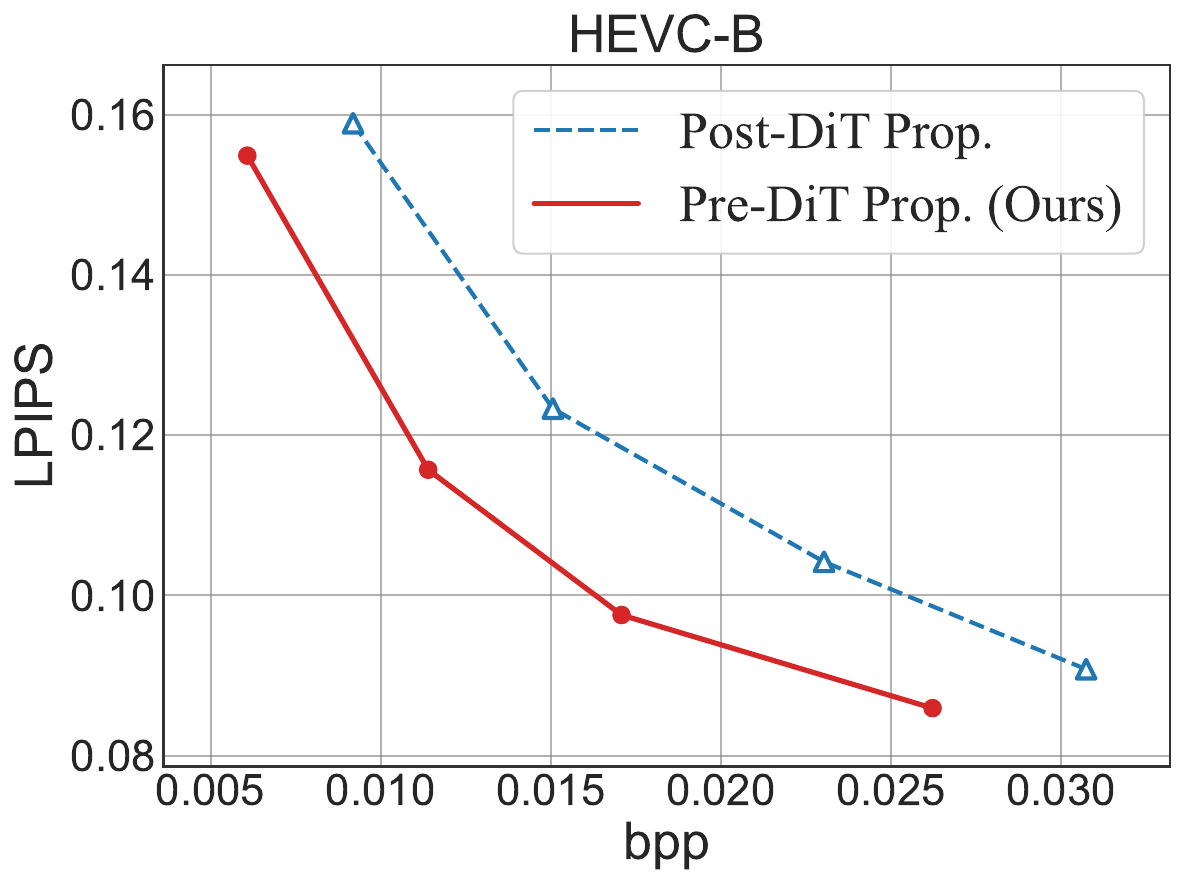}
    }
    \hfill 
    \subfloat[]{
        \label{fig:ablation_dists_prop}
        \includegraphics[width=0.47\linewidth]{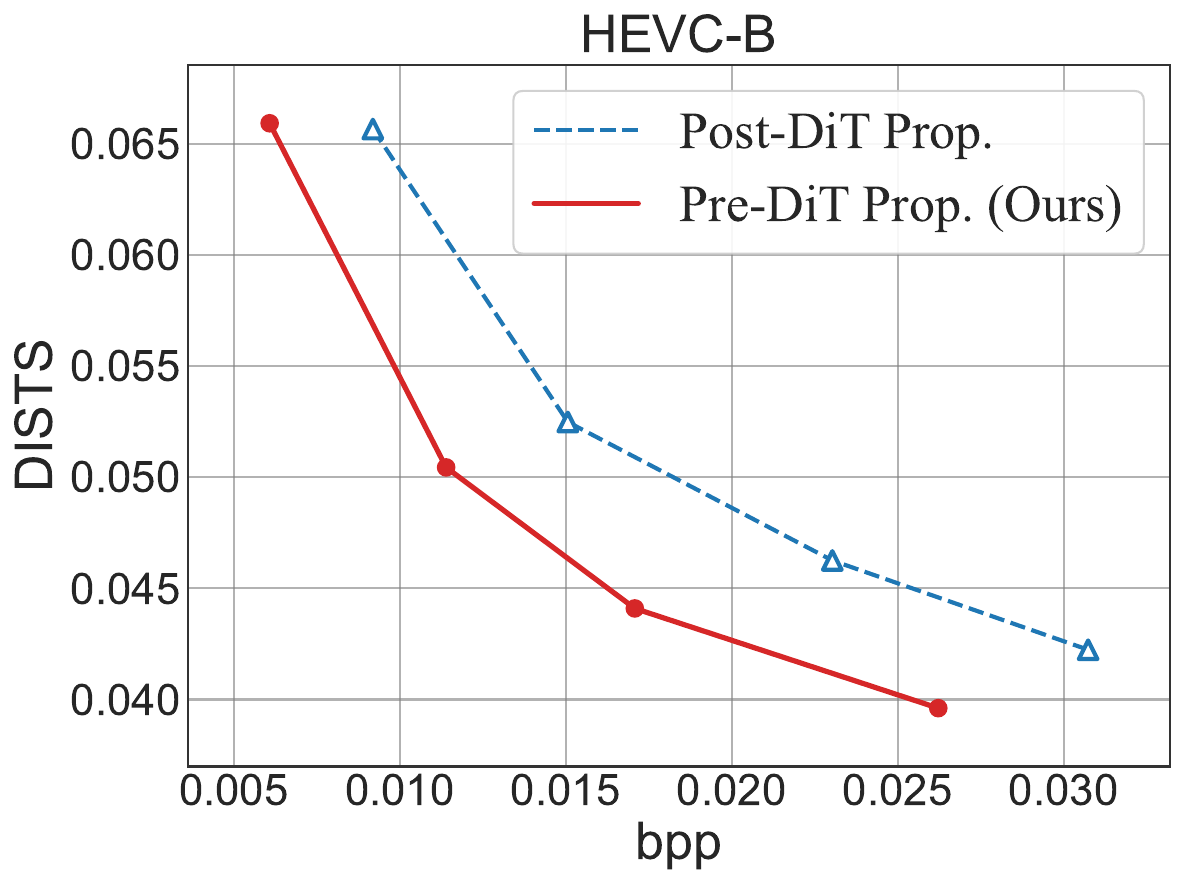}
    }
\caption{BD-rate comparison of temporal prior embedding using the \emph{Pre-DiT} and \emph{Post-DiT} strategies.} 
\label{fig:ablation}
\end{figure}

Currently, we directly cache the 256-channel feature \(\hat{f}_t\) from the CLC decoder before DiT (the \emph{Pre-DiT} strategy) and use it as the temporal condition (prior) to generate \(\hat{F}_t\) and \(\hat{F}_h\) in the CLC (see Fig.~\ref{fig:compare_framework2}). Since DiT further improves reconstruction quality, a natural question arises: can we instead use the DiT-denoised output (the \emph{Post-DiT} strategy) to generate \(\hat{F}_t\) and \(\hat{F}_h\) for conditional coding?

To investigate this alternative, we propose the Post-DiT variant illustrated in Fig.~\ref{fig:compare_framework1} and retrain the model for a fair comparison. The experimental results in Fig.~\ref{fig:ablation} show that the Post-DiT strategy actually underperforms the Pre-DiT design, contradicting the initial intuition. This degradation is likely due to the information loss incurred when compressing \(\hat{f}_t\) from 256 channels down to the 32-channel input required by DiT, which limits the usefulness of the DiT-denoised features to propagate sufficient temporal information.

\begin{figure}[htbp]
    \centering
    \subfloat[]{
    \label{fig:denoise_vis_a}
    \includegraphics[width=1\linewidth]{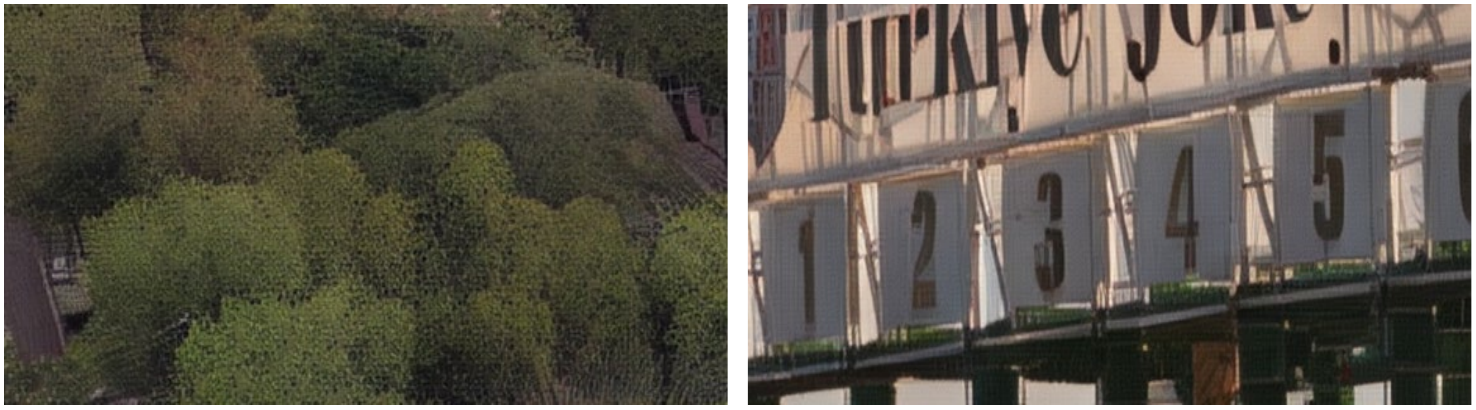}
    }
    \hfill 
    \subfloat[]{
    \label{fig:denoise_vis_b}
    \includegraphics[width=1\linewidth]{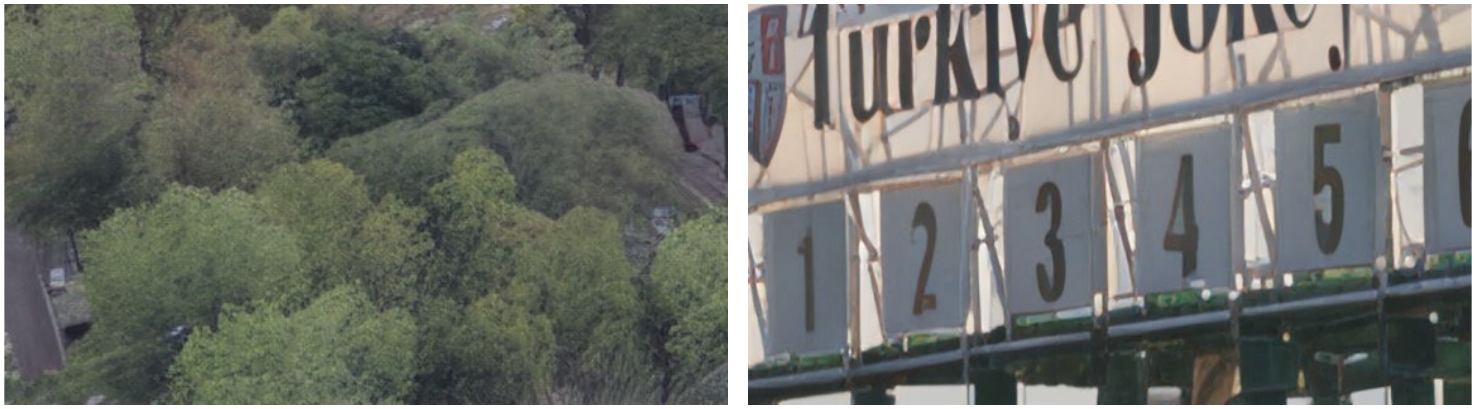}
    }
\caption{Visual comparison of decoded images via TA-AE using (a) the compressed latent \(\tilde{l}\) output directly from the CLC before DiT, and (b) the denoised latent \(\hat{l}\) produced by DiT.}
    \label{fig:denoise_vis}
\end{figure}

\subsubsection{One-Step DiT Denoiser} 
\begin{table}[htbp]
    \centering
    \caption{Performance-complexity tradeoff with or without DiT.}
    \label{tab:dit_ablation}
    \setlength{\tabcolsep}{4mm} 
    \resizebox{\linewidth}{!}{ 
    \begin{tabular}{l|cc|cc}
        \toprule
        \multirow{2}{*}{\textbf{Configuration}} 
            & \multicolumn{2}{c}{\bf Decoding Time} 
            & \multicolumn{2}{c}{\textbf{BD-Rate}} \\
        \cmidrule(lr){2-3} \cmidrule(lr){4-5}
         & 1080p & 480p & LPIPS $\downarrow$ & DISTS $\downarrow$ \\
        \midrule
        w/o DiT in P& $\approx$0.922 & $\approx$0.190s & +12.44\% & +15.92\% \\
        \midrule
        w/o DiT &  $\approx$0.919 & $\approx$0.188s & +14.93\% & +20.72\% \\
        \midrule
        \addlinespace
        \rowcolor{OursBlue}
        w/ DiT & $\approx$1.016s & $\approx$0.266s & 0.00\% & 0.00\% \\
        \bottomrule
    \end{tabular}}
\begin{flushleft}
        \footnotesize $^{\dagger}$Average time over 32 frames (1 I-frame and 31 P-frames).
    \end{flushleft}
\end{table}

To clearly demonstrate the role of the DiT, we visualize the decoded images both before and after the denoising stage. Specifically, we feed the CLC-decoded latent \(\tilde{l}_t\) and the denoised latent \(\hat{l}_t\) into the TA-decoder to generate the corresponding frames. As shown in Fig.~\ref{fig:denoise_vis}, the frame reconstructed from \(\tilde{l}_t\) exhibits noticeable artifacts, whereas the DiT-denoised output shows substantially improved visual quality, confirming that DiT acts as an effective latent-space denoiser.

To further quantify DiT’s contribution, we first remove it in P-frames only (see Fig.~\ref{fig:ipp}) and retrain the model, denoted as “w/o DiT in P”. We then remove DiT from both I-frames and P-frames, yielding a second variant denoted “w/o DiT”.

{In Table~\ref{tab:dit_ablation}, eliminating DiT in P-Frames already causes a substantial degradation in performance while yielding only marginal latency savings.  Removing DiT in I-Frame further reduces the performance. These results demonstrate that the DiT component is essential for achieving high performance, both quantitatively and qualitatively.}

%% file: sec/5_conclusion.tex
\section{Conclusion}

This paper presents YODA, a high-fidelity neural video compression framework that extends latent diffusion models to video by incorporating explicit temporal awareness and enabling efficient single-step inference. 
YODA uses a Temporal-Aware AutoEncoder (TA-AE) to embed multiscale temporal features from reference frames into latent representation learning; a Conditional Latent Coder (CLC) with channel expansion to propagate rich, high-dimensional context across frames; and a LoRA-finetuned linear Diffusion Transformer (DiT) for one-step latent denoising.
Extensive experiments on various public datasets demonstrate that YODA consistently achieves state-of-the-art perceptual performance, outperforming both traditional codecs (e.g., VTM) and recent neural approaches (e.g., PLVC and GLC-Video) across prevalent perceptual metrics like LPIPS, DISTS, FID, and KID. 

{\bf Limitations \& Future Directions.} YODA currently supports only low-delay encoding with an IPPP structure; future work may extend it to more flexible configurations with bidirectional prediction. Moreover, as indicated by the decoding latency in Table~\ref{tab:complexity}, real-time processing is not yet achievable with the present design. This makes further optimization of model architecture, inference efficiency, and hardware-aware implementations highly desirable for practical deployment in real-time and interactive video applications.

\section*{Acknowledgment}
We thank the authors of DC-AE~\cite{dcae}, DCVC-RT~\cite{DCVCRT}, SANA~\cite{sanasprint}, and related works for their pioneering contributions and open-source efforts. We will also release YODA publicly to facilitate further research and development.